\documentclass[aps,prl,twocolumn,showpacs,superscriptaddress,groupedaddress]{revtex4-2}
\usepackage{graphicx}% Include figure files
\usepackage{dcolumn}% Align table columns on decimal point
\usepackage{bm}% bold math
\usepackage{textcomp,gensymb}
\usepackage[T1,T2A]{fontenc}
\usepackage[utf8]{inputenc}
\usepackage{enumitem}
\usepackage{longtable}
\usepackage{tabu}
\usepackage{amsmath}
\usepackage{placeins}
\usepackage{natbib}
\usepackage{url}
\usepackage{amsfonts}
\usepackage{amssymb}
\usepackage{amsthm}
\usepackage{float}
\begin{document}
\preprint{APS/123-QED}
\title{Unraveling the temperature-dependent spin-polarized electron transport in iron via spin-wave Doppler shift}
%%%%%%%%%%%%%%%%%%
\author{J. Solano}
\email{jose.solano@fzu.cz}
\affiliation{Institute of Physics, Czech Academy of Sciences, Cukrovarnická 10, 162 00 Praha 6, Czech Republic}
%%%%%%%%%%%%%%%%%%
\author{Q. Rossi}
\author{J. Robert}
\author{M. Lenertz}
\author{Y. Henry}
\author{B. Gobaut}
\author{D. Halley}
\author{M. Bailleul}
\email{matthieu.bailleul@ipcms.unistra.fr}
\affiliation{Institut de Physique et Chimie des Matériaux de Strasbourg, UMR 7504 CNRS, Université de Strasbourg, 23 rue du Loess, BP 43, 67034 Strasbourg Cedex 2, France}

%%%%%%%%%%%%%%%%%%%%%%%%%%%%%%%%%%%%%%%%%%%%%%%%%%%%%%%%%%%%%%%%%%%%%%%%
\begin{abstract}

An electric current flowing in a ferromagnetic metal carries spin angular momentum, i.e. it is spin-polarized. Here, we measure the spin-wave Doppler shift induced by the transfer of angular momentum from the diffusive spin-polarized electric current onto coherent spin waves in epitaxial MgO/Fe/MgO thin films. We follow this Doppler shift as function of the temperature and determine that the degree of spin-polarization of the current increases from 77$\%$ to 86$\%$ when cooling the device from 303K down to 10K. Interpreting these measurements within the two-current model, we separate the contributions from electron-surface, electron-phonon and electron-magnon scatterings to the spin-dependent resistivity of Fe.

\end{abstract}
\maketitle

Spintronics incorporates the electron spin degree of freedom in information storage and processing by exploiting spin currents~\cite{Dieny.2020}. Part of this research deals with diffusive spin-polarized transport in metallic ferromagnets, such as Fe, Co, Ni and their alloys~\cite{Coey.2012}. This has been a constant research topic in the field since the pioneering work~\cite{Campbell.1967,Fert.1968,Loegel.1971} that led to the invention of the giant magnetoresistance (GMR). It continues to be of interest today, as this transport plays a critical role in the performance of metal-based devices such as GMR sensors~\cite{Reig.2013} and spin-torque nano-oscillators~\cite{Jiang.2024}.

\begin{figure}[ht]
\includegraphics[width=88mm]{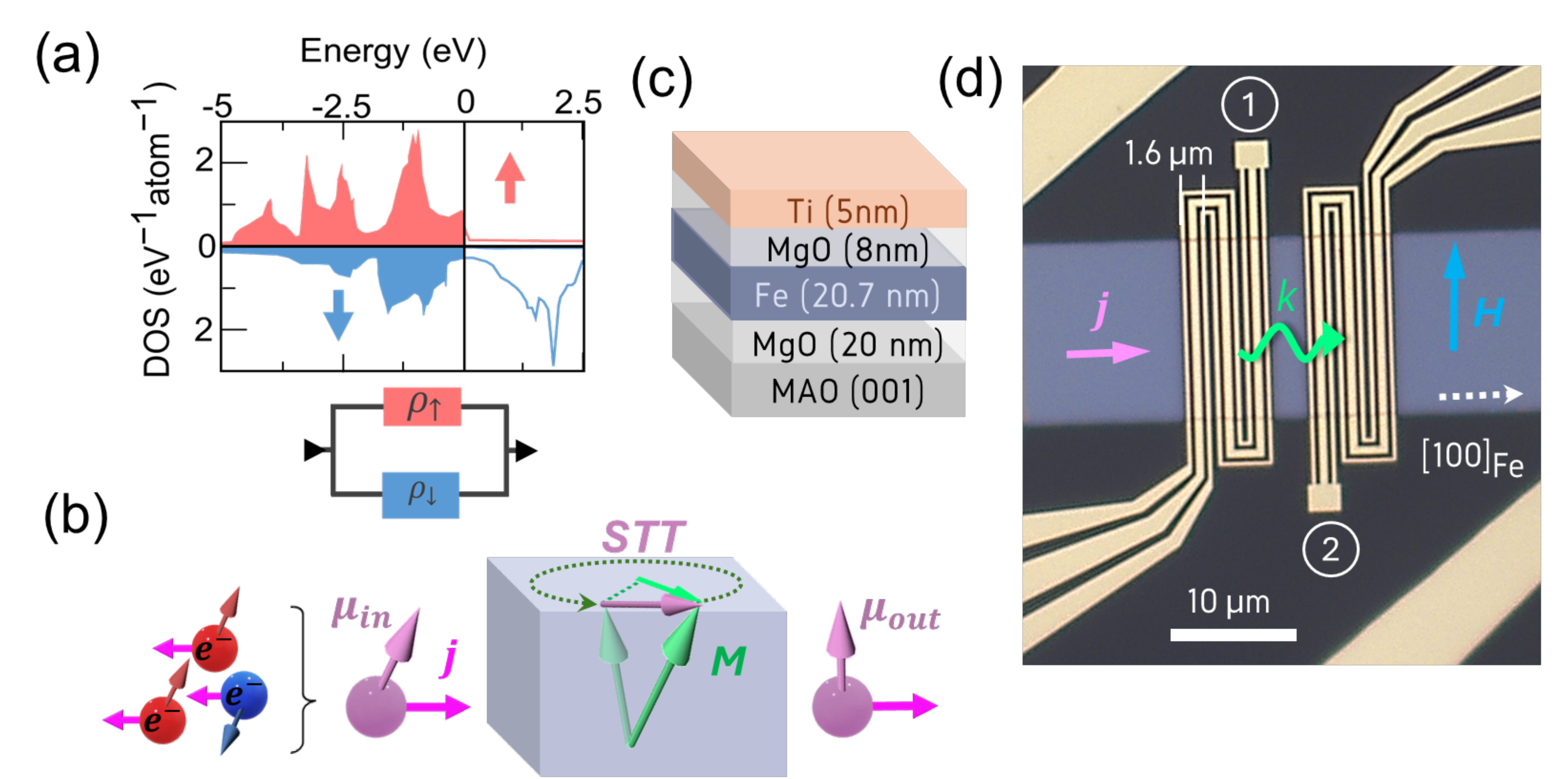}%
\caption{\label{fig:Intro} (a) Calculated density of states for bcc Fe for the two spin directions (data taken from Ref.~\cite{Coey.2012}) and sketch of the two-current model with two parallel channels of conduction. (b) Sketch of the adiabatic spin-transfer torque exerted on the spin wave magnetization by a spin-polarized electric current. (c) Sketch of the film stack. (d) Experimental configuration sketched on top of an optical microscopy photo of a nanofabricated device (substrates is black, stack strip is blue and microwave antennas are yellow).}
\end{figure}

Spin-polarized transport in ferromagnetic metals originates from the exchange-splitting of the spin $\uparrow$ and spin $\downarrow$ electron bands [Fig.~\ref{fig:Intro} (a)]. In the Boltzmann picture of transport, this exchange-splitting translates into different densities, velocities and scattering times for majority and minority Fermi-level electron states, resulting in different resistivities $\rho_{\uparrow}$ and $\rho_{\downarrow}$, respectively, as described in Mott’s two-current model~\cite{Mott.1964} [Fig.~\ref{fig:Intro} (a)]. To have an adequate understanding of this transport, one must account precisely for the spin-polarized electron bands and their interaction with defects (impurities, alloy disorder, interfaces), lattice displacements (phonons) and thermal spin-disorder (magnons). While bulk defect contributions have been extensively studied in the early days of spintronics, both theoretically~\cite{Mertig.1995} and experimentally~\cite{Campbell.1967,Fert.1968,Loegel.1971}, thermal contributions remain poorly understood. 
 
From the theoretical point of view, advanced ab inito calculations required for a realistic treatment of scattering by phonons and magnons were only made available recently~\cite{Ebert.2015,Verstraete.2013,Liu.2015}. Moreover, the role of electron correlations remains an open question~\cite{Nabok.2021}, particularly regarding electron-magnon coupling which is by itself a complex many-body physics effect~\cite{Paischer.2023}. From the experimental point of view, the major difficulty resides in the fact that the electrical resistivity (the only easily accessible observable) is a combination of contributions of the two spin states [Eq.~\eqref{eq:rho}]. 

\begin{equation}
 \rho = 
   \frac{\rho_\uparrow \rho_\downarrow }{
   \rho_\uparrow + \rho_\downarrow },\label{eq:rho}
\end{equation}
%%%%%%%%%%%%%%%%

Traditional techniques employed to separate these contributions (deviations from Mathiesens’s rule, giant magnetoresistance, point-contact Andreev reflection) are indirect; and limited to a very low temperature range, and/or material composition~\cite{Campbell.1982,Zhu.2008,Soulen.1998}. In this context, the so-called spin-wave Doppler shift method~\cite{Vlaminck.2008} is particularly useful as it gives direct access to the spin current carried by electrons, regardless of the temperature range. This method measures the frequency shift that coherent spin waves experience due to the spin-transfer torque from a spin-polarized electron current to the magnetization [Fig.~\ref{fig:Intro} (b)]. Under the two-current model, this frequency shift~\cite{Lederer.1966} is directly proportional to the degree of spin-polarization of the current $P$ [Eq.~\eqref{eq:polarization}].

%%%%%%%%%%%%%%%%
\begin{equation}
    P=\frac{\rho_\downarrow-\rho_\uparrow }{\rho_\uparrow + \rho_\downarrow}.\label{eq:polarization}
\end{equation}
%%%%%%%%%%%%%%%%

The technique has been successfully applied to polycrystalline Ni$_{80}$Fe$_{20}$ films, for which alloy disorder-dominated values of $P$ = 0.5-0.8 have been extracted~\cite{Vlaminck.2008, Sekiguchi.2012, Chauleau.2014, Haidar.2013, Zhu.2010, Haidar.2013, Liu.2015} and validated by comparison with other relevant measurements and calculations~\cite{Liu.2015}. More recently, this technique revealed that epitaxial Fe thin films have a very high $P=0.83$ at room temperature~\cite{Gladii.2016.thesis}. This is a surprising result since Fe, being a weak ferromagnet, is expected to have a relatively small spin-polarization of its density of states and of its ballistic electron current ($\sim$ 0.36-0.60)~\cite{Campbell.1982, Tsymbal.1996, Zhu.2008, Mazin.1999, Zhu.2008, Soulen.1998}. The disagreement regarding spin-polarization between different transport regimes remains unclear, as their comparison is challenging~\cite{Zhu.2008}. Additionally, ab initio simulations have not fully addressed the spin-polarization problem~\cite{Ebert.2015, Ma.2023}, as defining spin states is non-trivial in relativistic calculations, and there is no temperature-dependent experimental data contribute to these efforts.

In this work, we report and analyze temperature-dependent resistivity $\rho$ and degree of spin-polarization $P$ measurements for epitaxial Fe films. The measurement of these two quantities [Eqs.~\eqref{eq:rho} and~\eqref{eq:polarization}] allows us to separate resistivities of the two electron spin-channels. Having separated them, we use simple models to interpret these resistivities as a combination of electron-surface, electron-phonon, and electron-magnon scatterings.

% Fig1 %%%%%%%%%%%%%%%%%
%%%%%%%%%%%%%%%%

The MgO(20nm)/Fe(20.7nm)/MgO(8nm)/Ti(4.5nm) films [Fig.~\ref{fig:Intro} (c)] are grown on MgAl$_\text{2}$O$_\text{4}$ (001) substrates by molecular beam epitaxy. The cristallinity of the layers is monitored during growth by reflective high energy electron diffraction and after growth by X-Ray diffraction. The films are also characterized by X-Ray reflectometry, resistivity, magnetometry and ferromagnetic resonance (FMR)~\cite{Supplemental, Amorim.2021}. We obtain properties similar to those of films grown on MgO substrates~\cite{MagnifouetTchinda.2020, Solano.2022, Solano.2024, Devolder.2013}. Strong indications of film homogeneity are rocking curves with width $<1^{\circ}$, small FMR linewidth (damping $\alpha$=2.7$\times10^{-3}$, inhomogeneous broadening $\Delta H_0$=0.6 mT) and residual resistivity ratio $\rho_{300K}/\rho_{0K}\text{=6.6}$ \footnote{The room temperature resistivity is $\rho_{300K}\text{=11.8} \; \mu \Omega \text{cm} $ while the extrapolated value to 0 K is $\rho_{0K}\text{=1.8} \; \mu \Omega \text{cm} $ }. Additionally, from the FMR characterizations we obtain bulk Fe-like magnetic properties with well-defined surface anisotropies~\cite{Supplemental, Stringfellow.1968}.

The films undergo a nanofabrication process to pattern strips of ferromagnetic material that serve as conduits for the electric current and the propagating spin waves. Additional steps allow one to pattern four DC contacts, and a pair of microwave antennas on top of each of the strips for spin wave spectroscopy [see Fig.~\ref{fig:Intro} (d)]. The patterned devices are connected to a printed circuit board~\cite{Supplemental} inside a cryostat that is placed at the center of an electromagnet. The DC-pads are connected to a sourcemeter for 4-probe measurements while the antennas are connected to a 2-port microwave vector network analyzer (VNA).

\begin{figure}[ht]
\includegraphics[width=0.45\textwidth]{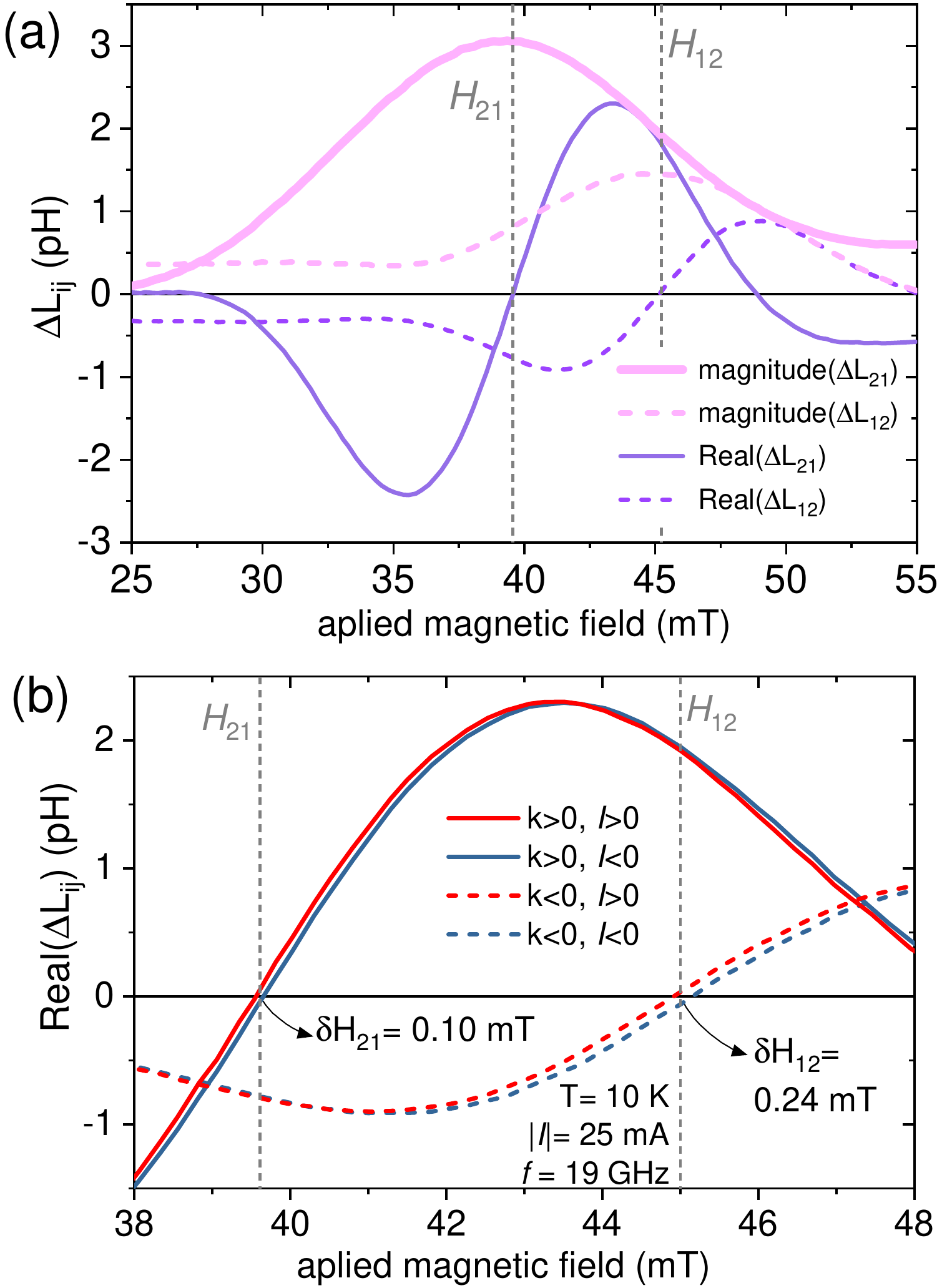}%
\caption{\label{fig:MutualInductance} Mutual inductance $\Delta L$ between antennas as function of the applied magnetic field at temperature $T$=10K and applied DC current (a) $|I|$=0 mA. (b) $|I|$=25 mA. The measured device consists of a strip with width 13.8 $\mu$m and edge-to-edge distance between antennas of 3 $\mu$m [see Fig.~\ref{fig:Intro} (a)].}
\end{figure}

Microwave measurements are performed at a fixed frequency of 19 GHz. The VNA measures the microwave scattering matrix of the connected device, from which we extract the mutual inductance between the two antennas $L_{ij}$, $i,j$=1,2. This is measured as a function of an external magnetic field $H$ applied in the plane of the film perpendicular to the spin-wave wave vector [so-called Damon Eschbach geometry, Fig.~\ref{fig:Intro} (d)]. Fig.~\ref{fig:MutualInductance} (a) displays the change of mutual-inductance $\Delta L_{ij}=L_{ij}(H)-L_{ij}(H_0)$ with respect to an out-of-resonance reference taken at a field $\mu_0 H_0=200\text{ mT}$. When $H$ is swept around the resonance field $H_{ij}$, the excitation microwave field of one antenna drives the precession of the magnetization (ferromagnetic resonance) resulting in coherent spin waves that propagate until the region below the second antenna where their stray field is detected inductively. Then, around the resonance field $H_{ij}$, this process leads to a change in the mutual inductance between the antennas as shown in Fig.~\ref{fig:MutualInductance} (a). $\Delta L_{ij}$ is in fact a complex quantity that oscillates as spin waves with slightly different wave vectors (different time of flight between antennas) are excited inside the bandwidth of the antenna~\cite{Vlaminck.2010}. In our devices, the antenna excitation is centered around $k=3.9 \;\text{rad}/\mu\text{m}$ ($\lambda=1.6 \mu \text{m}$)~\cite{Supplemental}. Note that we measure $\Delta L_{ij}$ for spin waves propagating with $k>0$ ($ij=21$) and $k<0$ ($ij=12$). In Fig.~\ref{fig:MutualInductance}, the spectra of the counterpropagating waves displays non-reciprocal signal amplitude and resonance field. These non-reciprocities are discussed elsewhere~\cite{Supplemental} since they are do not affect the analysis below.

% Fig2 %%%%%%%%%%%%%%%%%

In a second step, we monitor the spin wave propagation with the additional injection of an electric current $\mathbf{j}$ in the Fe strip. In Fig.~\ref{fig:MutualInductance} (b) we present the real part of the mutual inductance due to counterpropagating spin waves in the strip for both polarities of the electric current. We observe that the spin wave spectra shifts in field depending on both the current polarity ($j>0$, $j<0$) and the propagation direction ($k>0$, $k<0$). To estimate the current-induced field shift $\delta H_{ij}$, we center ourselves at each resonance field $H_{ij}$, measure the horizontal shifts between the spectra corresponding to $j>0$ and $j<0$, and average those within a $\pm 0.5$ mT window.

There are two effects that give rise to these current-induced shifts~\cite{Haidar.2013}: the spin transfer torque from the spin-polarized current to the magnetization (spin wave Doppler shift) and the presence of an additional Oersted field produced by the electron current. The combination of these contributions with different symmetries gives rise to the non-reciprocal nature of the shifts $\delta H_{ij}$ displayed in Fig.~\ref{fig:MutualInductance} (b).  Then, by symmetry arguments~\cite{Supplemental}, we can extract the field shift due to the spin wave Doppler effect $\delta H_\text{Dopp}^*(I) = \frac{\delta H_\text{21}- \delta H_\text{12}}{4} -\delta H^*_\text{NROe}$ \footnote{$^*$ denotes the Doppler shift for $j<0$ and $k>0$}, where $\delta H^*_\text{NROe}$ is the nonreciprocal contribution of the Oersted field~\cite{Haidar.2014,Supplemental}.

\begin{figure}[ht]
\includegraphics[width=90mm]{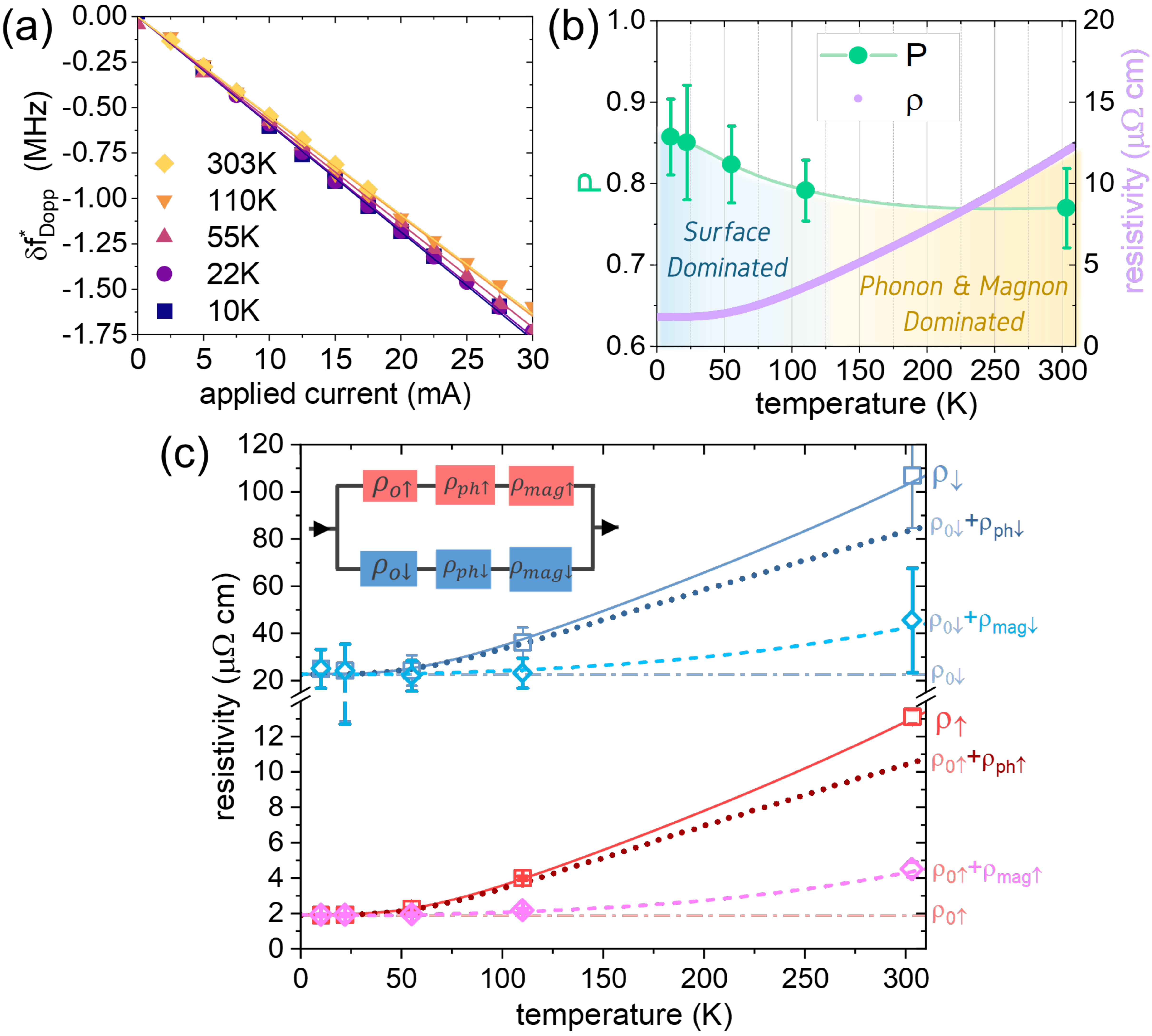}%
\caption{\label{fig:P&r} (a) Spin wave Doppler frequency shifts as a function of the applied DC current for different temperatures. The lines are the corresponding linear fits from which we extract the reported $P$ values. (b) Estimated degree of spin-polarization of the current (green squares) and measured resistivity (purple dots) as function of the temperature. The green line is a guide to the eyes. (c) Two-current model resistivity as a function of the temperature. Red and blue squares are the two spin channel resistivities that we decompose into the following contributions: the horizontal lines are the residual resistivities, the dotted lines are the adjustments to Eq.~\ref{eq:BlochWilson2}, the diamonds are calculated as $\rho_{\text{i}}-\rho_{ph,\text{i}}$ while the dashed lines are the respective fits to $B_i\rho^*_{mag}$ (see discussion below). The solid red and blue lines are the sums of the three contributions to spin up and spin down channels resistivity, respectively.}
\end{figure}

We can transform these field shifts \footnote{The current and temperature dependence of $\delta H_\text{Dopp}^*$ are given in the supplemental material.} into the corresponding frequency shifts $\delta f_\text{Dopp}^*$ using the resonance frequency expression ($\delta f= \frac{\partial f}{ \partial H} \delta H$)~\cite{Supplemental} and our temperature-dependent ferromagnetic resonance study~\cite{Supplemental}. In Fig.~\ref{fig:P&r} (a) we present the spin wave Doppler frequency shifts as function of the applied current for different temperatures. We can fit the current dependence of these shifts to the following expression~\cite{Vlaminck.2008}:

%%%%%%%%%%%%%%%%
\begin{equation}
\begin{aligned}
    \label{eq:DopplerShift}
    \delta f_\text{Dopp} =  - \frac{\mu_\text{B}}{ 2\pi e M_\text{s}} P \: \mathbf{j} \cdot \mathbf{k},
\end{aligned}
\end{equation}
%%%%%%%%%%%%%%%%

where $M_\text{s}$ is the saturation magnetization. From the slopes we extract the temperature dependence of $P$ that we report in Fig.~\ref{fig:P&r} (b) using $k=3.9 \; \text{rad/}\mu \text{m}$ and the measured temperature dependence of $M_\text{s}$~\cite{Supplemental}. 

From this analysis we obtain the main result of our investigation: the electric current in Fe is highly spin-polarized (87 $\%$ to 77$\%$) in the entire temperature range 10 K-303 K [Fig.\ref{fig:P&r} (b)] \footnote{We have also performed standard spin wave Doppler shift measurements by sweeping frequency instead of applied field~\cite{Vlaminck.2008,Haidar.2013,Gladii.2017} at room temperature from which we obtain the same value of $P$(300K)=0.77}. This confirms the room temperature result of Gladii et al.~\cite{Gladii.2016.thesis,Gladii.2017} suggesting a highly spin-polarizing effect of the electron scattering with thermal excitations (phonons and magnons). Furthermore, when this scattering becomes negligible at low temperatures, $P$ increases slightly. We attribute this low temperature high spin-polarization to scattering with the film's surfaces.  

The transition between thermal and surface scattering sources becomes clearer when looking at the resistivity $\rho$. 
In this case, we observe an important increase above the residual resistivity only at 110 K, when the thermally induced resistivity becomes comparable to the residual one [Fig.\ref{fig:P&r} (b)]. This motivates our separation of the electron scattering into a surface dominated regime ($T\lesssim$110K), where electron-surface scattering is the main contribution; and a thermal regime ($T\gtrsim$110K) where scattering with phonons and magnons dominates.

% Fig3 %%%%%%%%%%%%%%%%%

Now that we have estimated the spin-polarization of the current and measured the resistivity as a function of the temperature, we proceed to analyze the data further under the two-current model of Eqs.~\eqref{eq:rho} and ~\eqref{eq:polarization}. In Fig.\ref{fig:P&r} (c) we plot the separate resistivities of the two channels of conduction $\rho_\uparrow$ and $\rho_\downarrow$. Here we observe the natural consequence of the high spin-polarization: the current is mainly carried by the spin-up electrons ($\rho_\uparrow \approx \rho$), while the spin-down conduction is highly resistive ($\rho_\downarrow \gg \rho$). Note the interrupted $y$-axis, with a scaling factor of ten between the two parts. We continue the study of the electron transport by separating the resistivity contributions to each spin channel originating from three scattering sources: surfaces, phonons and magnons.

\textit{Surfaces.}\textbf{——} In the absence of impurities and grain boundaries in our films, the only remaining source of scattering at low temperatures ($\sim$ 10 K) are the interfaces with MgO which we assume to exhibit some roughness at the atomic scale. The resulting spin-polarized residual resistivities are  $\rho_{\text{o},\uparrow}=1.95 \; \mu \Omega \text{cm}$ and $\rho_{\text{o},\downarrow}=23 \; \mu \Omega \text{cm}$.

We argue that this highly spin-polarized surface scattering could originate from the interplay between the Fe and MgO electron band structures proposed in previous studies~\cite{Bowen.2001,Butler.2001}. Our hypothesis is that interface atomic roughness induces less scattering for spin-up electrons than for spin-down electrons resulting in the high $P$ we observe. Indeed, spin-up electron wave functions of the so-called $\Gamma_1$ symmetry display long-range evanescence in the MgO layer~\cite{Butler.2001}, then they are expected to be less sensitive to the interface atomic scale roughness. In the vocabulary of the Fuchs-Sondheimer model~\cite{Sondheimer.1952}, this translates into a smaller specularity coefficient for the majority electrons~\cite{Sondheimer.1952, Zhou.2018, Supplemental}. From the analysis of our spin-wave measurements, we believe that most of this surface scattering occurs at the top Fe/MgO interface, which contrary to the bottom one is not annealed. Indeed, this interface exhibits a smaller perpendicular magnetic anisotropy~\cite{Supplemental,Solano.2022}, which is associated to a lower degree of ordering~\cite{Hallal.2013}. Moreover, the measured current-induced Oersted shift indicates that the electrical current distribution is not symmetric across the strip's cross section, with a larger amplitude towards the bottom of the film~\cite{Solano.2024, Supplemental} .
    
\textit{Phonons.}\textbf{——} To model the resistivity due to electron-phonon scattering we consider a spin-polarized version of the Bloch-Wilson model where phonons scatter highly conductive electrons into heavier states of the same spin direction (so called s-d scattering)~\cite{Wilson.1938}. The spin-polarized resistivity expressions are given in Eq.~\eqref{eq:BlochWilson2}:

%%%%%%%%%%%%%%%%
\begin{equation}
\begin{aligned}
    \label{eq:BlochWilson2}
   \rho_{ph,\text{i}} = A_i \left( \frac{T}{\Theta_\text{D} }\right)^3 \int_{0}^{\Theta_\text{D} /T} \frac{z^3 dz}{(e^z-1)(1-e^{-z})},
\end{aligned}
\end{equation}
%%%%%%%%%%%%%%%% 

where $\Theta_\text{D}\text{=477 K}$ is the Debye temperature for bulk iron~\cite{Tari.2003} and $A_i$ are spin dependent resistivity amplitudes that we proceed to adjust now to our data.

We assume that most of the temperature-dependent resistivity of each spin-channel is given by electron-phonon scattering: we adjust $A_i$ so that this component accounts for most of $\rho_i$ while keeping $\rho_{\text{o},i}+\rho_{\text{ph},i}<\rho_i$ [Fig.~\ref{fig:P&r} (c)]. This strong assumption is inspired by ab initio calculations in refs.~\cite{Verstraete.2013,Ebert.2015, Ma.2023}, where this type of scattering accounts for up to 70$\%$ of Fe's resistivity. Moreover, below we present independent magnetoresistance measurements that corroborate this hypothesis.  

Our adjusted $\rho_{ph \uparrow}$ and $\rho_{ph \downarrow}$ are larger than the ones deduced from first principle calculations~\cite{Verstraete.2013, Ma.2023}, by a factor of 1.5 and 5, respectively. Consequently, our estimate of electron-phonon scattering has a higher spin-polarization $P_\text{ph}=(\rho_{ph \downarrow}-\rho_{ph \uparrow})/(\rho_{ph \downarrow}+\rho_{ph \uparrow}) = 0.75$ than the calculated one $P_\text{ph} \approx 0.33$~\cite{Verstraete.2013}. This disagreement may be attributed to the absence of spin-orbit coupling (SOC) and correlations in the calculations. Indeed, Ma et al. have shown that, accounting for SOC significantly increases the total resistivity due to electron-phonon scattering while it also flattens the minority electron bands close to the Fermi level~\cite{Ma.2023}. Additionally, similar modifications to the minority electron bands have been obtained by including nonlocal correlations due to electron-magnon interactions \cite{Paischer.2023}. Then, our hypothesis is that these effects are responsible for reducing the dispersion of the spin-down electron bands, hence the large $\rho_{ph \downarrow}$ and high $P$ we observe. Meanwhile, the highly dispersive spin-up band is not modified importantly~\cite{Ma.2023, Paischer.2023}, and our estimated $\rho_{ph \uparrow}$ remains close to the simulated value without SOC in ref.~\cite{Verstraete.2013,Ma.2023}.

\textit{Magnons.}\textbf{——} Now we can look at the remaining components of the resistivities: $\rho_{mag,\text{i}}=\rho_{\text{i}}-\rho_{\text{o},i}-\rho_{ph,\text{i}}$. Their temperature dependence shows a pronounced curvature [Fig.~\ref{fig:P&r} (c)] that can be reproduced neither by the Bloch-Wilson model nor by ab initio simulations of the electron-phonon scattering~\cite{Verstraete.2013,Ebert.2015,Ma.2023}. Then, we attribute these remaining resistivities to electron-magnon scattering: conductive electron states of a given spin band are scattered into heavier states of opposite spin after spin-flip collisions with magnons.

To model these contributions we resort to the theory first developed by Goodings and later expanded by Raquet et al. that predicts the temperature dependent resistivity $\rho^*_{mag}(T)$ \footnote{We use * to denote the resistivity resulting from the non spin-polarized electron-magnon scattering model. See ref.~\cite{Goodings.1963}} arising from the s-d electron-magnon scattering~\cite{Goodings.1963,Raquet.2002}. As we did in the case of the phonon contribution, we modify the model by including spin-polarized renormalization factors $B_i$ to the temperature-dependent resistivity: $\rho_{mag,i}=B_i\rho^*_{mag}$. In doing so, we obtain $B_\uparrow=1.3$ and $B_\downarrow=10.9$ [see Fig.~\ref{fig:P&r} (c)]. This large spin-polarization should originate from the large probability of spin-flip scattering of spin-down conduction electrons into the very state dense/flat spin-up band crossing the Fermi level. Contrarily, spin-up conduction electrons should have lower probability of spin-flip scattering even after modifications by SOC or correlations due to the lower density of states of the spin-down bands.

Note that we have only considered spin-flip scattering events in which a conduction electron is scattered by a magnon into a heavy state of opposite spin ($\rho_{mag,i}$). A second type of events involve spin-flip scattering only between conduction electrons, resulting in a resistivity contribution due to current mixing ($\rho_{\uparrow\downarrow}$)~\cite{Fert.1969}. We estimate this second contribution using the analytical model developed in ref.~\cite{Fert.1969} for low temperatures and find it is negligible~\cite{Supplemental}. We argue that the very high and almost constant spin-polarization we estimate is an indication that the dominant spin-flip scattering events are those that contribute to the polarized resistivity of the spin channels ($\rho_{mag,i}$) rather than to their current mixing ($\rho_{\uparrow\downarrow}$)~\cite{Supplemental}. This suggests that in Fe the primary role of spin-flip-inducing interactions, such as SOC, may not be to decrease $P$ through spin-mixing, as is often assumed \cite{Verstraete.2013}. Instead, these interactions may even enhance $P$ by reducing the dispersion of the spin-down electron bands as suggested above.

Finally, as an independent check of our model and its hypotheses, we resort to an alternative experimental method to estimate the electron-magnon resistivity $\rho^*_{mag}$~\footnotemark[\value{footnote}] from the so-called magnon magnetoresistance effect~\cite{Raquet.2002}. For this purpose we measure the longitudinal magnetoresistance $\Delta \rho(H,T) = \rho(H,T)-\rho(0,T)$ at high magnetic field and high enough temperatures~\cite{Raquet.2002}. This method assumes that the main effect of a strong magnetic field is to shift the magnon dispersion to higher energies (depopulating the thermal magnon bath) effectively decreasing electron-magnon scattering and the associated resistivity. By measuring such magnon-magnetoresistance over a broad enough range of field and temperature, one has enough information to reconstruct the temperature dependence of $\rho^*_{mag}(T)$ at a given magnetic field as shown in ref.~\cite{Raquet.2002}.

\begin{figure}[ht]
\includegraphics[width=85mm]{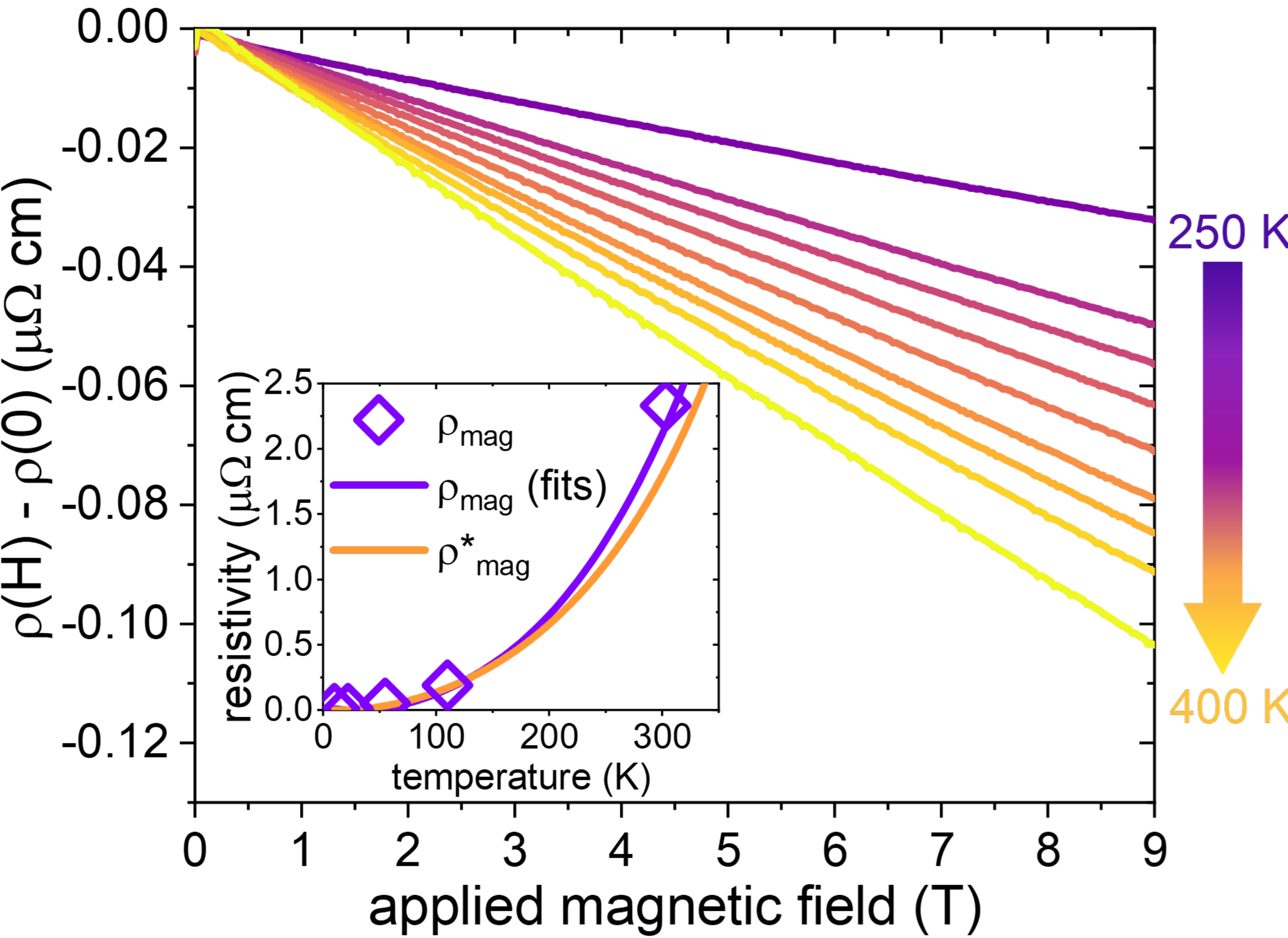}%
\caption{\label{fig:MMR} Longitudinal magnetoresistance as function of applied magnetic field for different temperatures. The inset shows the estimated total resistivity due to electron-magnon scattering: obtained from our two-current model $\rho_{mag}$ (diamonds), fits to $B_i\rho^*_{mag}$ (purple line) and the reconstruction $\rho^*_{mag}$ from our longitudinal magnetoresistance measurements for an applied field $\mu_0 H$=120 mT (orange line).}
\end{figure}

In Fig.~\ref{fig:MMR} we present the longitudinal magnetoresistance measurements in the same devices that we used for the Doppler measurements. In this case, we apply the external magnetic field parallel to the current (along Fe's [100] easy axis) and we measure the longitudinal voltage to obtain the resistance. We observe a close-to-linear reduction of $\rho$ as a function of the magnetic field with a gradually steeper decrease as temperature increases. We fit the magnetoresistance to the model proposed by Raquet et al.~\cite{Raquet.2002} and obtain the spin wave stiffness $D(T)=356[1-(2\times10^{-6})T^2]$ $\text{meV}\text{\AA}^2$~\cite{Supplemental} which is in good agreement with their result. We reconstruct the electron-magnon resistivity $\rho^*_{mag}$ following the methods of ref.~\cite{Raquet.2002} and plot it as the orange line in the inset of Fig.~\ref{fig:MMR}.

% Fig4 %%%%%%%%%%%%%%%%%

Now, we can compare this result to the one obtained previously in our two-current model [Fig.~\ref{fig:P&r} (c)]: $\rho_{mag}=\left(\rho_{mag\uparrow}\:\rho_{mag\downarrow}\right)/\left(\rho_{mag\uparrow}+\rho_{mag\downarrow}\right)$ \footnote{here we use the fact that the spin-polarization of the different process are all high (about 0.8), such that overall resistivities can be added (no major deviation from Matthiessen's rule}. In the inset of Fig.~\ref{fig:MMR} we can see the excellent agreement between the two experimental methods. As the two estimates of the electron-magnon resistivity originate from two independent measurements probing two different regimes (low vs. high magnetic field and temperature), we believe that this agreement is a strong indication that our two-current model is a good overall representation of the spin-polarized electron transport in Fe.

%%%%%%%%%%%%%%%%%%%%%%%%%%%%%%%%%%%%%%%%%%%%%%%%%%%%%%%%%%%%%%%%%%%%%%%%%%%%%%%%%%%%%%%%
%Conclusion
%%%%%%%%%%%%%%%%%%%%%%%%%%%%%%%%%%%%%%%%%%%%%%%%%%%%%%%%%%%%%%%%%%%%%%%%%%%%%%%%%%%%%%%%%
To conclude, by accessing the temperature dependence of the spin-polarization of the current in Fe we have shown that majority electrons are much less affected by scattering and carry most (roughly 80$\%$) of the current over the entire 10-300K temperature range. This temperature dependence allowed us to further separate the contributions from films surfaces, phonons and thermal magnons to the resistivity in a relatively simple model. These results suggest that these three mechanisms have similar degrees of spin-polarization, which is surprising in the conventional picture of electron transport, considering the different regions of Fermi surfaces and electron scattering potential at play for these processes. As we emphasize, the observed electron transport exhibits greater spin polarization than predicted by traditional density of states considerations, prompting the need to account for non-conventional effects in the analysis. We hope our results will motivate the extension and refinement of existing simulations of electrical transport in itinerant ferromagnets, addressing directly spin-polarized transport in the presence of thermal excitations of both the crystal lattice and the magnetic order and possibly also surface scattering effects. We anticipate spin-polarized transport could be a relevant observable to address finite-temperature non-colinear-spins electron physics and explore the influence of spin-orbit coupling and electron-electron correlations \cite{Paischer.2023, Stepanov.2022}.

\begin{acknowledgments}
We thank Sergiy Mankovsky, Benjamin Bacq-Labreuil and François Gautier for theoretical support and insightful discussion; Romain Bernard, Sabine Siegwald and Hicham Majjad for technical support during nanofabrication at STnano platform, Arnaud Boulard, Benoît Leconte, Daniel Spor, Jérémy Thoraval and Fares Abiza for their support in assembling and testing the broadband FMR and PSWS setup. We acknowledge financial support by the Interdisciplinary Thematic Institute QMat, as part of the ITI 2021-2028 program of the University of Strasbourg, CNRS and Inserm, IdEx Unistra (ANR 10 IDEX 0002), SWING project part of PEPR SPIN, SFRI STRAT’US project (ANR 20 SFRI 0012) and ANR-17-EURE-0024 under the framework of the French Investments for the Future Program. We also acknowledge financial support from Region Grand Est through its FRCR call (NanoTeraHertz and RaNGE projects) and from Agence Nationale de la Recherche (France) under Contract No. ANR-20-CE24-0012 (MARIN).
\end{acknowledgments}
%%%%%%%%%%%%%%%%%%%%%%%%%%%%%%%%%%%%%%%%%%%%%%%%%%%%%%%%%%%%%%%%%%%%%%%%%%%%%%%%%%%%%%%%%

\bibliography{Fe_SPTransport}% Produces the bibliography via BibTeX.
\bibliographystyle{apsrev4-2}
\end{document}

\preprint{APS/123-QED}
\title{Supplemental material: \\
Unraveling the temperature dependent spin-polarized electron transport in epitaxial Fe films via spin wave Doppler shift}
%%%%%%%%%%%%%%%%%%

\maketitle

%%%%%%%%%%%%%%%%%%%%%%%%%%%%%%%%%%%%%%%%%%%%%%%%%%%%%%%%%%%%%%%%%%%%%%%%%%%%%%%%%%%%%%%
\section{\label{sec:SamplePreparation}Sample preparation\protect}

\subsection{Growth}

The MgO(20nm)/Fe(20nm)/MgO(8nm)/Ti(4.5nm) films were grown on Codex single-side polished 10$\times$10 mm MgAl$_\text{2}$O$_\text{4}$ (001) substrates by molecular beam epitaxy. The substrates were cleaned in baths of acetone and IPA. Then, they were introduced in the high vacuum chamber (base pressure of 5$\times10^{-9}$ mbar). After, the substrate is subjected to a sequence of annealing steps reaching a maximum temperature of 600 $\text{\textdegree}$C, a MgO buffer layer of 20 nm was grown at 520 $\text{\textdegree}$C in order to obtain a fresh epitaxial surface for the next ferromagnetic layer. The 20 nm Fe layer is grown at room temperature, and subsequently annealed at 460 $\text{\textdegree}$C during 1.5 hours to promote recrystallization of the Fe layer. Finally, after cooling down the sample to room temperature the top MgO 8nm-thick layer is deposited, followed by a 4.5nm-thick Ti layer. The good cristallinity of the layers was monitored during growth by reflective high energy electron diffraction and after growth by X-Ray diffraction with a quality comparable to that reported before \cite{MagnifouetTchinda.2020,Solano.2022}. Using X-ray reflectometry we measure an actual thickness of $t$=20.7 nm for the Fe layer.

%%%%%%%%%%%%%%%%%%%%%%%%%%%%%%%%%
\subsection{Device Fabrication}

\begin{figure}[ht]
\includegraphics[width=0.85\textwidth]{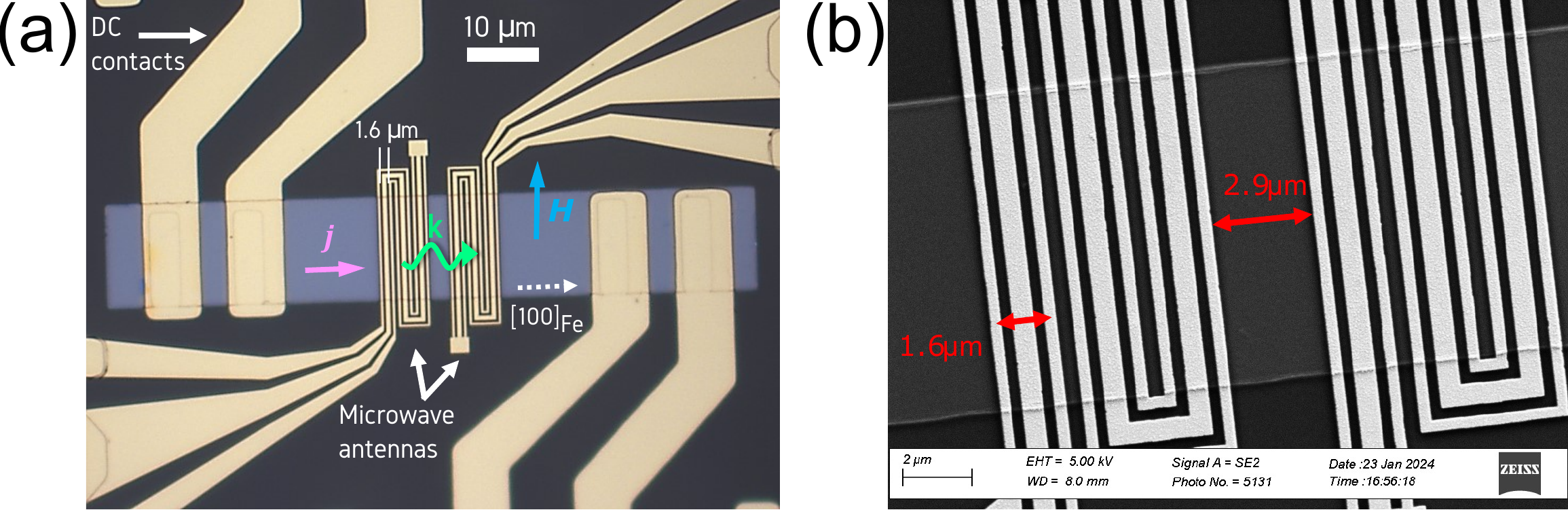}%
\caption{\label{fig:Device} (a) Sketch of the experimental configuration on top of an optical microscopy image of a nanofabricated device: 15$\mu$m width strip connected to 4 DC-probes with microwave antennas on top. (b) Scanning electron microscope image of the region between the two antennas.}
\end{figure}

The devices used consist of a strip of ferromagnetic material that serves as conduit for the electric current and propagating spin waves, of 4-probe DC contacts and of a pair of microwave antennas for spin wave spectroscopy (Fig.~\ref{fig:Device}). 

% Fig1 %%%%%%%%%%%%%%%%%
%%%%%%%%%%%%%%%%

1) To obtain useful devices we perform patterning of a strip (long edge along the [100] direction of the Fe lattice) by means of optical lithography (used as soft mask) and Ar$^+$ etching. Then, we sputter a 70 nm SiO$_2$ layer over the entire surface that serves as an insulating layer between the strip and the antennas, and also as protection layer for the MgO/MgAl$_\text{2}$O$_\text{4}$ exposed surface.

2) A second step of optical lithography and reactive ion etching is performed to open 4 adequate windows (4 $\mu$m$\times$13$\mu$m) in the SiO$_2$ for the electrical contacts of the next step.

3) A third optical lithography step is used as mask for Ar$^+$ etching to reach the Fe layer (etching of capping Ti and MgO) and the same mask is used for the deposition and lift-off of Ti(10nm)/Au(60nm) that form the 4-probe electrical contacts. During the etching step, we use a SIMS detector to reach roughly the middle thickness of the Fe layer. Note that during this step, we also etch the SiO$_2$ at the regions where the Au contacts will be connected by wire bond. This prevents cracking/detachment of the contact pads.

4) Finally, e-beam lithography is performed to pattern the microwave antennas that consist of a Ti(5nm)/Au(150nm) metalization. In Fig.~\ref{fig:Device} (b) we present a scanning electron microscope picture of the antennas.

%%%%%%%%%%%%%%%%%%%%%%%%%%%%%%%%%
\subsection{Device Connection}

\begin{figure}[ht]
\includegraphics[width=150mm]{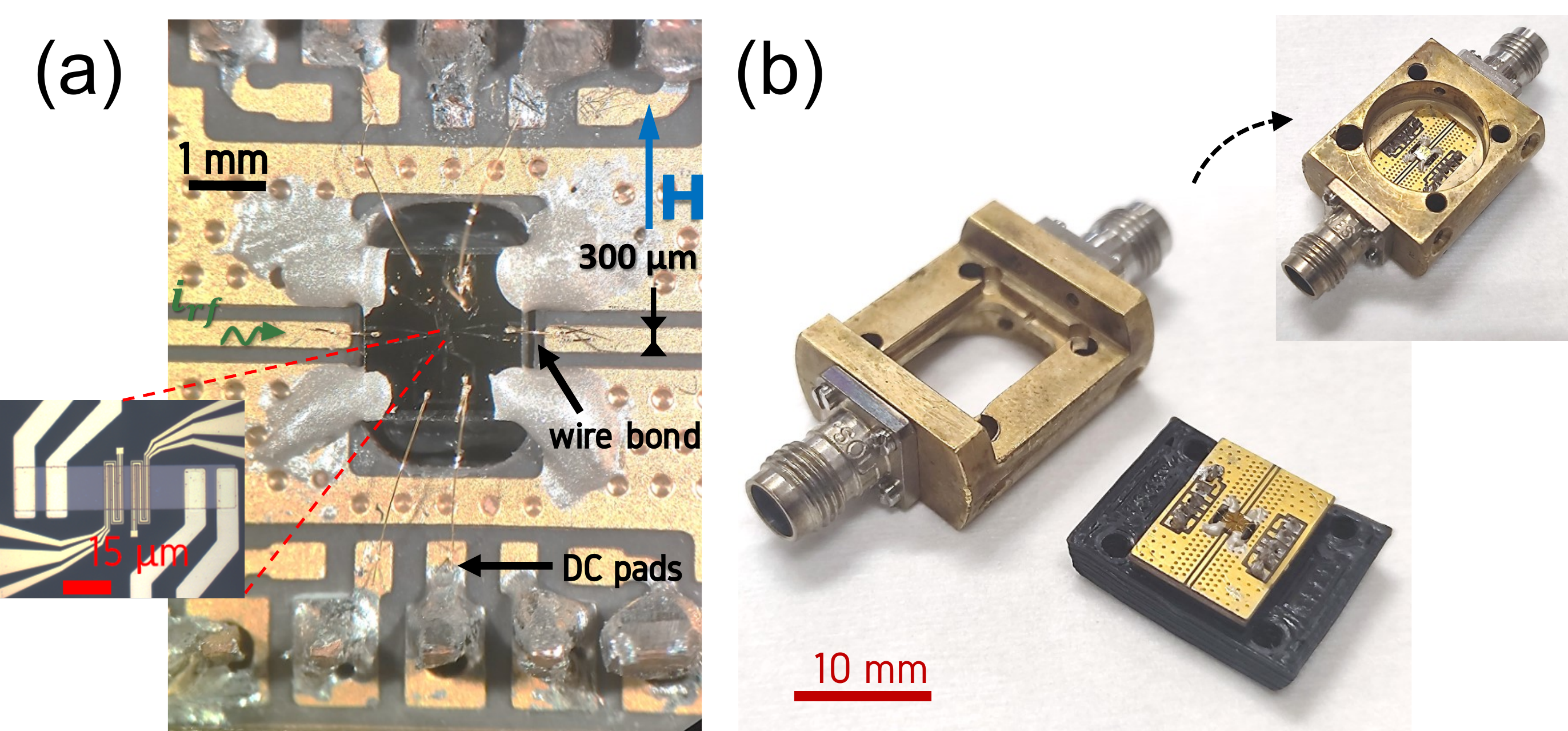}%
\caption{\label{fig:Box} (a) Optical microscopy photo of a sample (nanofabricated device) connected to a measuring PCB. (b) Photo of the complete sample-PCB interconnection and of the microwave box (2.4 mm connectors at each side) used for insertion in the cryostat.}
\end{figure}

To be able to measure the spin wave Doppler shift as a function of temperature, we connect the fabricated devices to a custom printed board circuit (PCB). This has been designed in-house and its coplanar waveguide has similar properties to the ones reported in Ref.~\cite{Solano.2022}. The main difference consists of the presence of pads for DC connections, and a cavity at the center for placing the nanofabricated device that is to be connected to the PCB [Fig.~\ref{fig:Box} (a)]. 

% Fig2 %%%%%%%%%%%%%%%%%
%%%%%%%%%%%%%%%%

The center lines of the sample's CPW (antennas) are connected to the center line of the PCB through Au wire bonds (diameter of 25 $\mu$m). On the other hand, the ground planes are interconnected with a conductive silver epoxy CW2400. Note that in the sample region there is no back copper and the sample is supported by a 3D printed plastic piece [black piece in Fig.~\ref{fig:Box} (b)]. This interconnection geometry guarantees a good RF transmission between the two CPWs (device and PCB) and prevents the feeding of spurious modes into the sample's substrate. 

The PCB is inserted in the helium atmosphere of a sample-in vapor cryogen-free ARS cryostat with a temperature range 5-400 K. The portion of the cryostat that contains the PCB is at the center of an electromagnet that provides the external magnetic field $H$. The DC pads of the PCB are connected to a sourcemeter for the transport measurements, while the two antennas are connected to a two-port vector network analyzer through 2.4 mm connectors and coaxial microwave cables.

%%%%%%%%%%%%%%%%%%%%%%%%%%%%%%%%%%%%%%%%%%%%%%%%%%%%%%%%%%%%%%%%%%%%%%%%%%%%%%%%%%%%%%%
\section{\label{sec:characterization}Experimental characterization\protect}

%%%%%%%%%%%%%%%%%%%%%%%%%%%%%%%%%
\subsection{SQUID magnetometry}

To measure the saturation magnetization of the films we used a Quantum Design MPMS3 SQUID magnetometer. The 10x10 $\text{mm}^2$ samples were cut into 2x2 $\text{mm}^2$ or 4x4 $\text{mm}^2$ pieces and then glued to a quartz holder with transparent nail polish. During the measurements, the magnetic field $H$ is applied parallel to the film's plane and along Fe's [100] easy axis.

To accurately determine the magnetic moment $\mu(H)$, we used a protocol developed by Amorin et al.\cite{Amorim.2021} that provides a correction factor $\alpha$ for the measurement of the magnetic moment. This correction factor is determined from the measurements themselves by exploiting the relationship between the moment values determined from the DC and VSM modes of the magnetometer. Post measurement, diamagnetic contributions were corrected by determining the diamagnetic susceptibility at high fields and subtracting it from the hysteresis loops. Finally, we used the film thickness $t$ (measured by X-ray reflection) and the measurement of the surface area $S$ of the films by optical microscopy to estimate the magnetization: 

%%%%%%%%%%%%%%%%
\begin{equation}
\begin{aligned}
    \label{eq:magnetizationVsH}
    M(H)= \frac{1}{\alpha} \frac{\mu(H)}{t S}.
\end{aligned}
\end{equation}
%%%%%%%%%%%%%%%% 

\begin{figure}[H]
\centering
\includegraphics[width=0.85\textwidth]{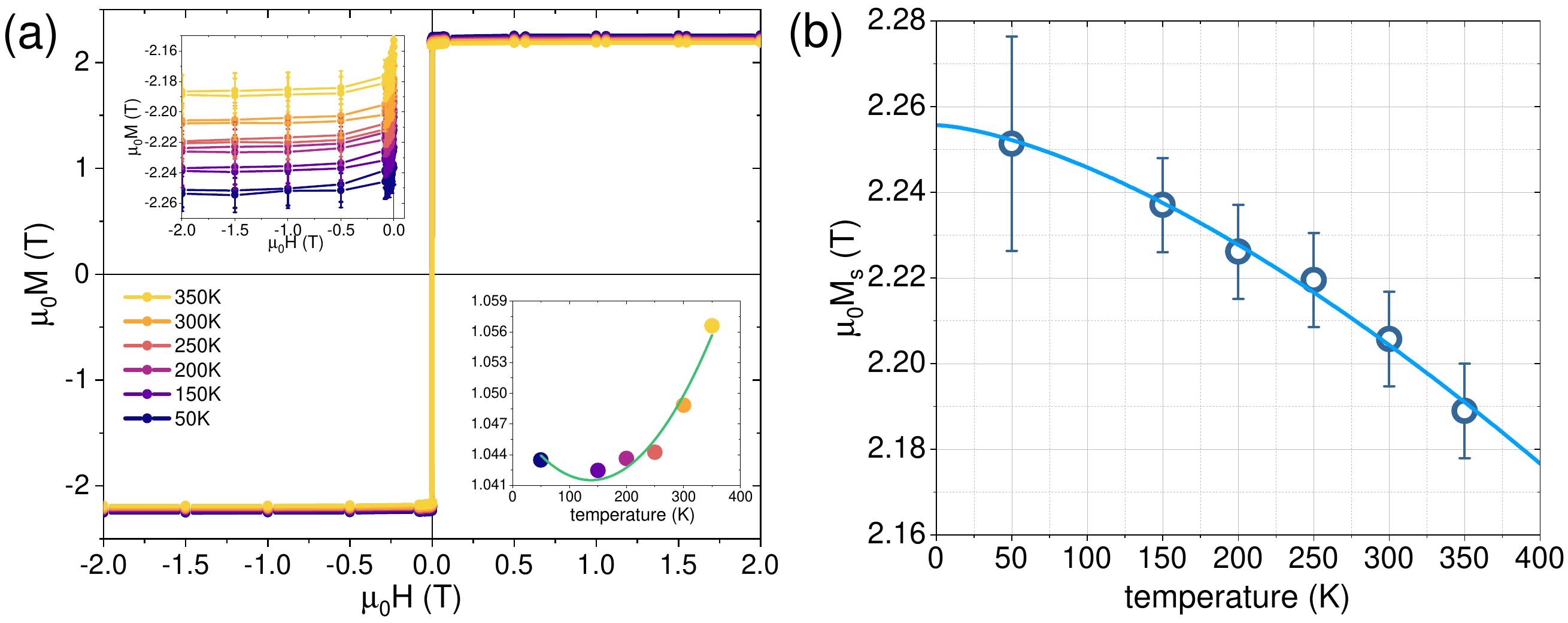}
\caption{(a) Temperature dependent magnetization hysteresis loops of the Fe films ($H$ field applied along in-plane easy axis). The top inset shows a zoom into the saturation region. The bottom inset shows the estimated temperature dependence of the correction factor. The green line is a guide to the eyes. (b) Temperature dependence of the saturation magnetization of our Fe films obtained from the hysteresis loops. The blue line is the fit to Bloch's law.} 
\label{fig:MvsT}
\end{figure}

To measure accurately the temperature evolution of the magnetization, we systematically determined the correction factor $\alpha(T)$ for each temperature [see inset of Fig.~\ref{fig:MvsT} (a)].
    
From the corrected hysteresis loops [Fig.~\ref{fig:MvsT} (a)] we obtain the saturation magnetization $M_\text{s}$ ($\mu_0 H> 2.5 \; \text{T}$) [Fig.~\ref{fig:MvsT} (b)]. In Fig.~\ref{fig:MvsT}, we present the resulting temperature dependence of $M_\text{s}$ for the films. We fitted the data to Bloch's law, resulting in: $M_\text{s}(T)= M_0[1-(4.4\times 10^{-6})T^{3/2}]$, where $\mu_0M_0=2.25 : T$. 

%%%%%%%%%%%%%%%%%%%%%%%%%%%%%%%%%
\subsection{Ferromagnetic resonance}

We conducted our ferromagnetic resonance characterization using a protocol similar to the one presented in ref.~\cite{Solano.2022}. The main difference is that we inserted our sample (2$\times$2 mm$^2$ film piece) and measuring PCB in a cryostat to vary the temperature. Unfortunately, this does not allow us to fully calibrate the microwave network as a function of the temperature forcing us to proceed in a semi-uncalibrated state: we measure precisely the electrical delays from each port to the sample as a function of temperature and we correct them manually in their respective VNA port extensions. We cannot calibrate the losses in the network and therefore neither the amplitude of the signal. 

\begin{figure}[H]
\centering
\includegraphics[width=0.85\textwidth]{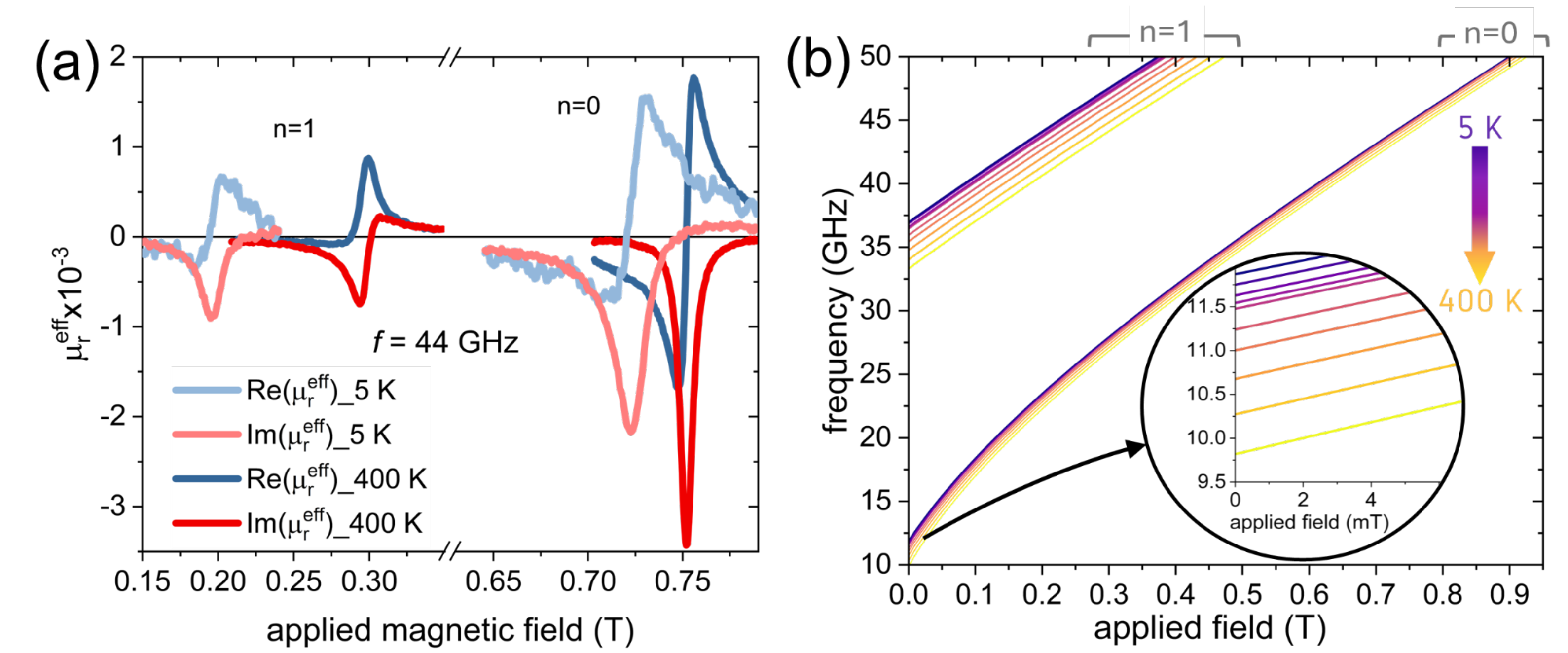}
\caption{(a) Ferromagnetic resonance for a thin film Fe (20.65nm) measured at different temperatures for in-plane applied magnetic field. (a) Resonance spectra at 44 GHz for 5 K and 400 K displaying two resonance peaks for each temperature, one for the homogeneous mode (n=0) and a second for the first standing spin wave mode (n=1). (b) Fits to Eq.~\eqref{eq:Kittel} for the resonance frequency as a function of the applied magnetic field for several temperatures in the range 5-400 K. We display the fits rather than the data for clarity purposes.} 
\label{fig:FMR}
\end{figure}

In the measurements we fix the microwave excitation at a given frequency (nominally at a power of -15 dBm), and sweep the applied magnetic field around the resonance condition. We apply the magnetic field in-plane (along [100] Fe's easy axis). Despite the not fully calibrated state of the network, this method allow us to determine accurately the resonance fields [Fig.~\ref{fig:FMR} (a)]. We followed the field dependence of the resonance frequency of the uniform ($n=0$) and first standing spin wave modes ($n=1$) for different temperatures [Fig.~\ref{fig:FMR} (b)].

Applying the model developed in ref.\cite{Solano.2022} for each temperature, we fit the resonance frequency as a function of the applied field to the expression:

%%%%%%%%%%%%%%%%
\begin{equation}
\begin{aligned}
    \label{eq:Kittel}
    f_{n}=\frac{\gamma\mu_0}{2\pi}[(H+H_{Xn})(H+H_{Yn})]^{1/2},
\end{aligned}
\end{equation}
%%%%%%%%%%%%%%%%

\begin{figure}[H]
\centering
\includegraphics[width=0.85\textwidth]{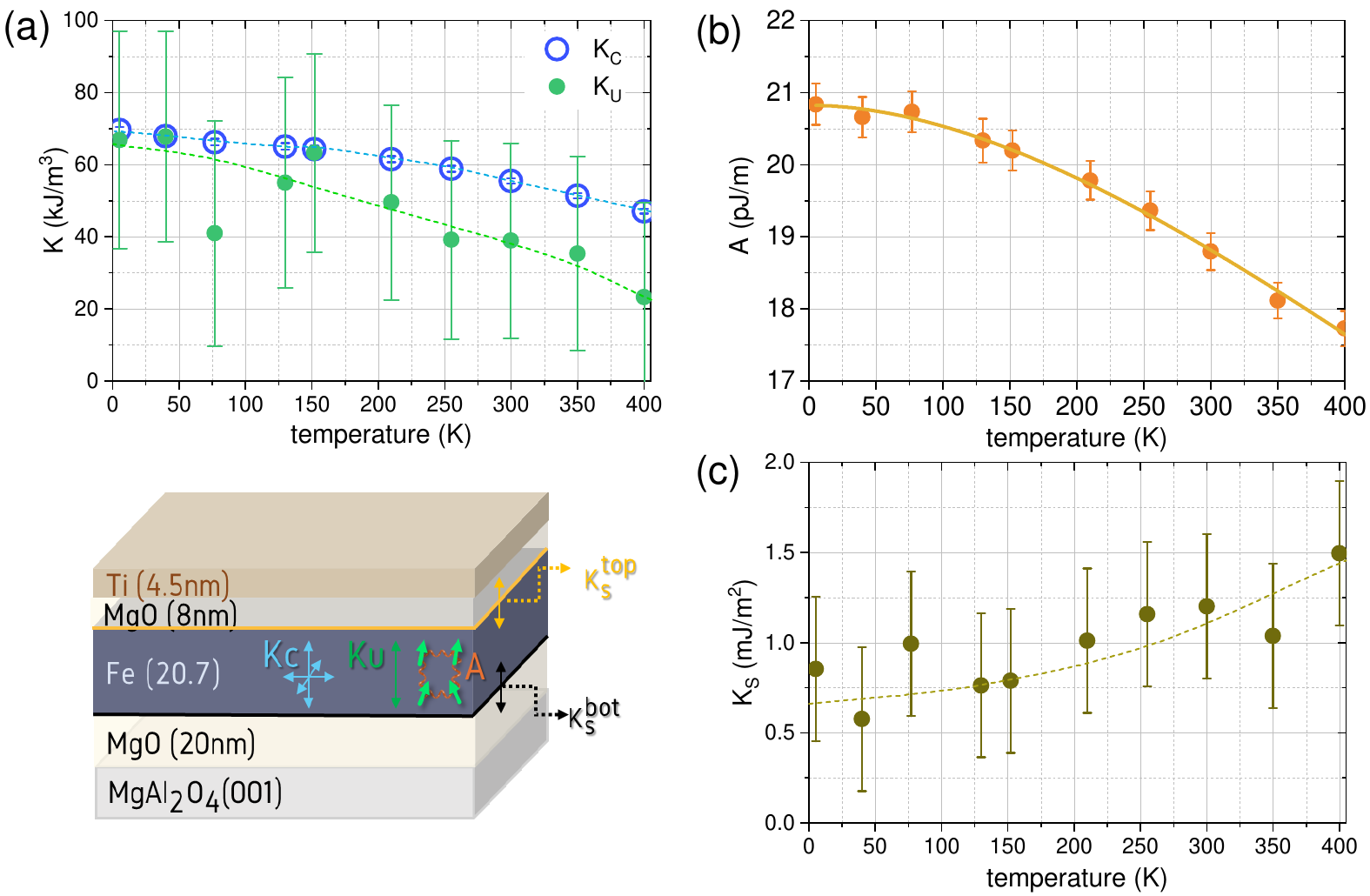}
\caption{Estimated magnetic parameters as a function of the temperature. (a) Cubic and volume uniaxial anisotropy constants (b) Exchange stiffness constant (c) Perpendicular uniaxial surface anisotropy. The dashed lines are just guides to the eyes. The continuous line for $A$ is a fit to a renormalization law \cite{Stringfellow.1968} resulting in: $A=20.83[1-1.8\times10^{-6}T^2+4\times10^{-8}T^{5/2}]$ pJ/m.} 
\label{fig:Parameters}
\end{figure}

From a symmetry analysis of the stiffness fields $H_{Xn}$ and $H_{Yn}$ we can extract the magnetic parameters for each explored temperature. The definitions of the stiffness fields for bcc Fe can be found in ref.~\cite{Solano.2022}. In Fig.~\ref{fig:Parameters} we present the resulting temperature dependence of the cubic anisotropy, exchange stiffness, volume uniaxial anisotropy (out-of-plane easy axis) and total perpendicular surface anisotropy in our films. For further details on the parameters and discussion of this characterization see refs. \cite{Solano.2022,Solano.2024}. We remark that the gyromagnetic ratio remains roughly constant over the entire temperature range with a value of $\gamma/2\pi=28.8\pm0.1$ GHz/T. In summary, this characterization shows that our films have a bulk Fe-like behavior of its magnetic parameters with some additional temperature dependent features that seem to be related to strain originating from the difference in elastic constants between the Fe film, the MgO layers, and the MgAl$_2$O$_4$ substrate. Using this temperature dependent characterization together with Eq.~\eqref{eq:Kittel} we can reconstruct the function $\partial f/ \partial H$ necessary to transform the field shifts into frequency shifts for the spin wave Doppler shift experiments.

%%%%%%%%%%%%%%%%%%%%%%%%%%%%%%%%%
\subsection{Propagating spin wave spectroscopy}

As in the case of FMR, the impossibility of calibrating our microwave network as a function of the temperature forced us to proceed in a semi-uncalibrated state: we measure precisely the electrical delay from each port to each antenna as a function of the temperature and we correct for them manually in their respective VNA port extensions. This allows us to measure the proper time delay due to the propagation of the spin waves between the antennas only. 

\begin{figure}[H]
\centering
\includegraphics[width=0.85\textwidth]{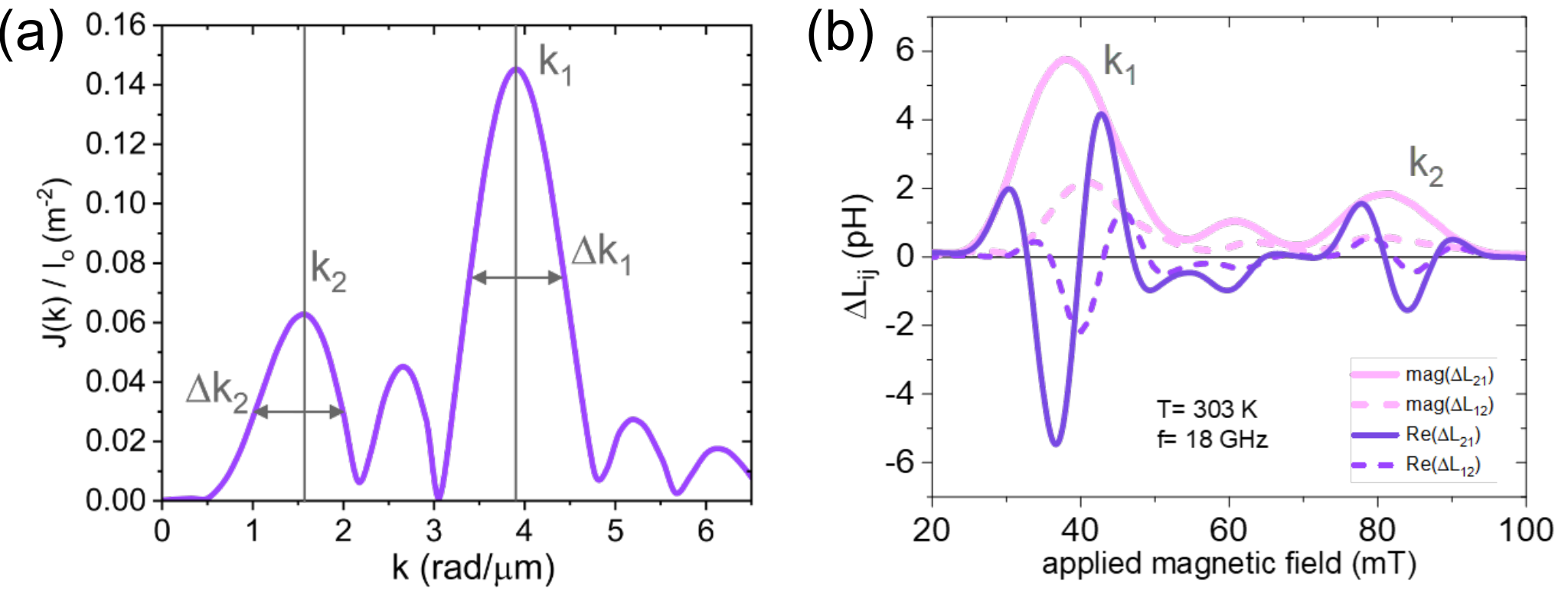}
\caption{(a) Calculation of the Fourier transform of the current density of an antenna displaying maxima around the wave vectors $k_1$ = 3.9 rad/$\mu$m and $k_2$ = 1.6 rad/$\mu$m assuming a uniform current density in the strands of the antennas. (b) Measured mutual inductance $\Delta L$ between the antennas as a function of the applied magnetic field at a temperature $T$=303K and excitation frequency 19 GHz. The edge-to-edge distance between the antennas is 2.9 $\mu$m and the Fe strip has a thickness of $t$=20.7 nm and a width of $w$=15 $\mu$m.} 
\label{fig:Excitation}
\end{figure}

Due to the semi-uncalibrated state, for the measurements we fix the microwave excitation at a given frequency (nominally 19 GHz at a power of -18 dBm), and sweep the applied magnetic field around the resonance condition. Note that this is opposite to the usual approach in which one fixes the applied magnetic field and sweep the frequency \cite{Gladii.2016, Solano.2022}. Although the geometry of our antennas produces two main excitation peaks around the wave numbers $k_1$=3.9 rad/$\mu$m and $k_2$=1.6 rad/$\mu$m (Fig.~\ref{fig:Excitation}), in this study we have focused on the propagation properties of the higher intensity excitation $k_1$ as it has the best signal to noise ratio.

\begin{figure}[H]
\centering
\includegraphics[width=0.85\textwidth]{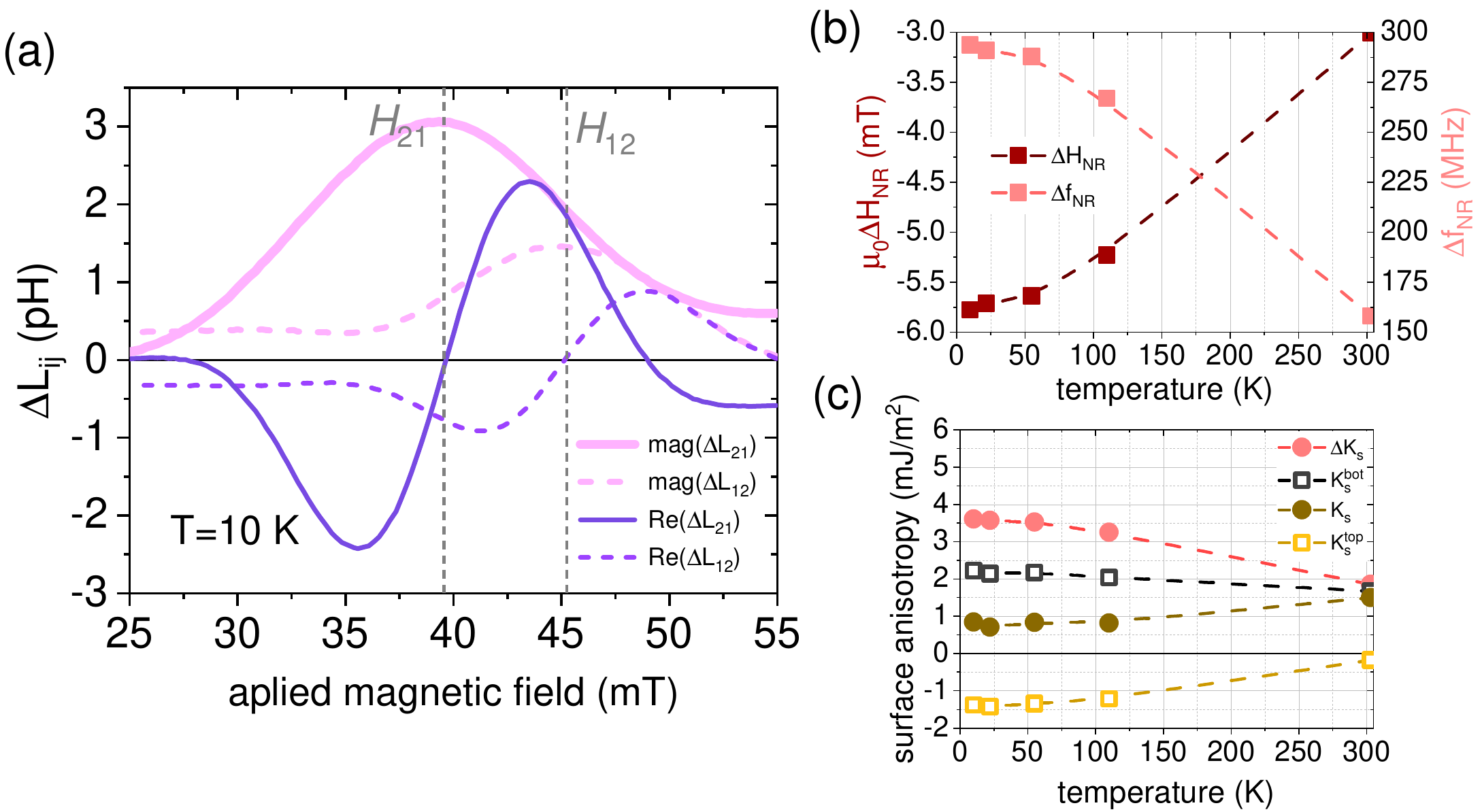}
\caption{(a) Mutual inductance $\Delta L$ between the antennas as a function of the applied magnetic field at a temperature $T$=10K. The signal corresponds to a wave number $k_1$=3.9 rad/$\mu$m and excitation frequency 18 GHz. The edge-to-edge distance between the antennas is 2.9 $\mu$m, the Fe strip has a thickness of $t$=20.7 nm and a width of $w$=15 $\mu$m (b) The measured field non-reciprocity and the corresponding calculated frequency non-reciprocity as a function of the temperature. (c) Estimated $\Delta K_\textbf{S}=K^\text{bot}_\text{S}-K^\text{top}_\text{S}$ and $K_\textbf{S}=K^\text{bot}_\text{S}+K^\text{top}_\text{S}$ and their separated components as a function of the temperature.} 
\label{fig:DKs}
\end{figure}

We observe important non-reciprocities between the spin wave signals of counter propagating waves. The first one is an important amplitude non-reciprocity that is well known for Damon–Eschbach waves \cite{Gladii.2016}. The second is a field non-reciprocity of the spin wave propagation, i.e. $\Delta H_{NR}=H_{k>0}-H_{k<0} \neq 0$ [see Fig.~\ref{fig:DKs} (a)]. This is associated to a frequency nonreciprocity $\Delta f_{NR}=f_{k<0}-f_{k>0}$ that we can calculate using the expression of the resonance frequency for Damon-Eshbach spin waves given in ref. \cite{Solano.2022}. In Fig.~\ref{fig:DKs} (b) we present the temperature dependence of these nonreciprocities. It has been shown that frequency non-reciprocity can be produced by a difference in perpendicular surface anisotropy between the bottom and top surfaces of the film, i.e. $\Delta K_\textbf{S}=K^\text{bot}_\text{S}-K^\text{top}_\text{S}\neq 0$ \cite{Gladii.2016} (see sketch in Fig.\ref{fig:Parameters}). From the propagation measurements, we estimate $\Delta K_\textbf{S}$ as a function of the temperature and using the temperature dependent $K_\textbf{S}=K^\text{bot}_\text{S}+K^\text{top}_\text{S}$ obtained from the ferromagnetic measurements above, we separate the contributions from the two interfaces of the film [Fig.~\ref{fig:DKs} (c)]. The value obtained for the bottom interface $K^\text{bot}_\text{S}$ match those expected for ideal Fe/MgO interfaces \cite{Hallal.2013}, while the negative value for the top interface $K^\text{top}_\text{S}$ suggests the presence of disorder \cite{Hallal.2013}.

%%%%%%%%%%%%%%%%%%%%%%%%%%%%%%%%%
\subsection{Spin wave Doppler shifts analysis}

As we mention in the main text, there are several contributions to the current-induced field shifts we observe experimentally. Our main interest is the contribution associated with the adiabatic spin-transfer torque process, the so called spin wave Doppler shift: electrons transfer angular momentum to the propagating spin waves changing their resonance frequency and their resonance field ($\delta H_\text{Dopp}$). A second contribution appears due to presence of an additional Oersted field $\delta H_\text{Oe}$ created by the applied electric current in the strip \cite{Haidar.2013}. Finally, there is a third non-reciprocal contribution $\delta H_\text{NROe}$ due to the effect of the $\delta H_\text{Oe}$ field on the inhomogeneous thickness-profile of the spin waves in the Damon-Eshbach geometry \cite{Haidar.2014}. For convenience, we define the following field contributions denoted by $^*$ to keep track of the sign of the wave vector $k$ and current $I$ in an abbreviated notation:

%%%%%%%%%%%%%%%%
\begin{subequations}
    \label{eq:ConventionFieldShifts}
     \begin{align}
     \delta H_\text{Dopp}^*(I) &= \delta H_\text{Dopp}(k>0,I<0), \label{eq:ConventionDopFieldShift}\\
     \delta H_\text{Oe}^*(I) &= \delta H_\text{Oe}(I<0), \label{eq:ConventionOeFieldShift}\\
     \delta H_\text{NROe}^*(I) &= \delta H_\text{Oe}(k>0,I<0).\label{eq:ConventionNROeFieldShift}
    \end{align}
\end{subequations}
%%%%%%%%%%%%%%%%

Due to the several contributions present in the current-induced field shifts, we need to combine different measurements in order to extract only $\delta H_\text{Dopp}^*$. Then, we can write the experimental field shifts for spin waves with $k>0$ as:

%%%%%%%%%%%%%%%%
\begin{equation}
\begin{aligned}
    \label{eq:21HShift}
   \delta H_\text{21}=& H_{21}(-I)-H_{21}(+I)\\
   =& H_{21}(I=0)+\delta H^*_\text{Dopp}+\delta H^*_\text{Oe}+\delta H^*_\text{NROe} - \left[H_{21}(I=0)-\delta H^*_\text{Dopp}-\delta H^*_\text{Oe}-\delta H^*_\text{NROe} \right]\\
   =&2\left(\delta H^*_\text{Dopp}+\delta H^*_\text{Oe}+\delta H^*_\text{NROe}\right),
\end{aligned}
\end{equation}
%%%%%%%%%%%%%%%% 

and for spin waves with $k<0$:

%%%%%%%%%%%%%%%%
\begin{equation}
\begin{aligned}
    \label{eq:DoppFieldShift}
   \delta H_\text{12}=& H_{12}(-I)-H_{12}(+I)\\
   =& H_{12}(I=0)-\delta H^*_\text{Dopp}+\delta H^*_\text{Oe}-\delta H^*_\text{NROe} - \left[H_{12}(I=0)+\delta H^*_\text{Dopp}-\delta H^*_\text{Oe}+\delta H^*_\text{NROe} \right]\\
   =&2\left(-\delta H^*_\text{Dopp}+\delta H^*_\text{Oe}-\delta H^*_\text{NROe}\right).
\end{aligned}
\end{equation}
%%%%%%%%%%%%%%%%

Finally, combining the shifts from counterpropagating waves we obtain:
%%%%%%%%%%%%%%%%
\begin{subequations}
    \label{eq:FieldShifts}
     \begin{align}
     \delta H_\text{Dopp}^*(I) + \delta H^*_\text{NROe}(I)&= \frac{\delta H_\text{21}- \delta H_\text{12}}{4} , \label{eq:DopFieldShift}\\
     \delta H_\text{Oe}^*(I) &= \frac{\delta H_\text{21}+ \delta H_\text{12}}{4}, \label{eq:OeFieldShift}\\
    \end{align}
\end{subequations}
%%%%%%%%%%%%%%%%

\begin{figure}[H]
\centering
\includegraphics[width=0.85\textwidth]{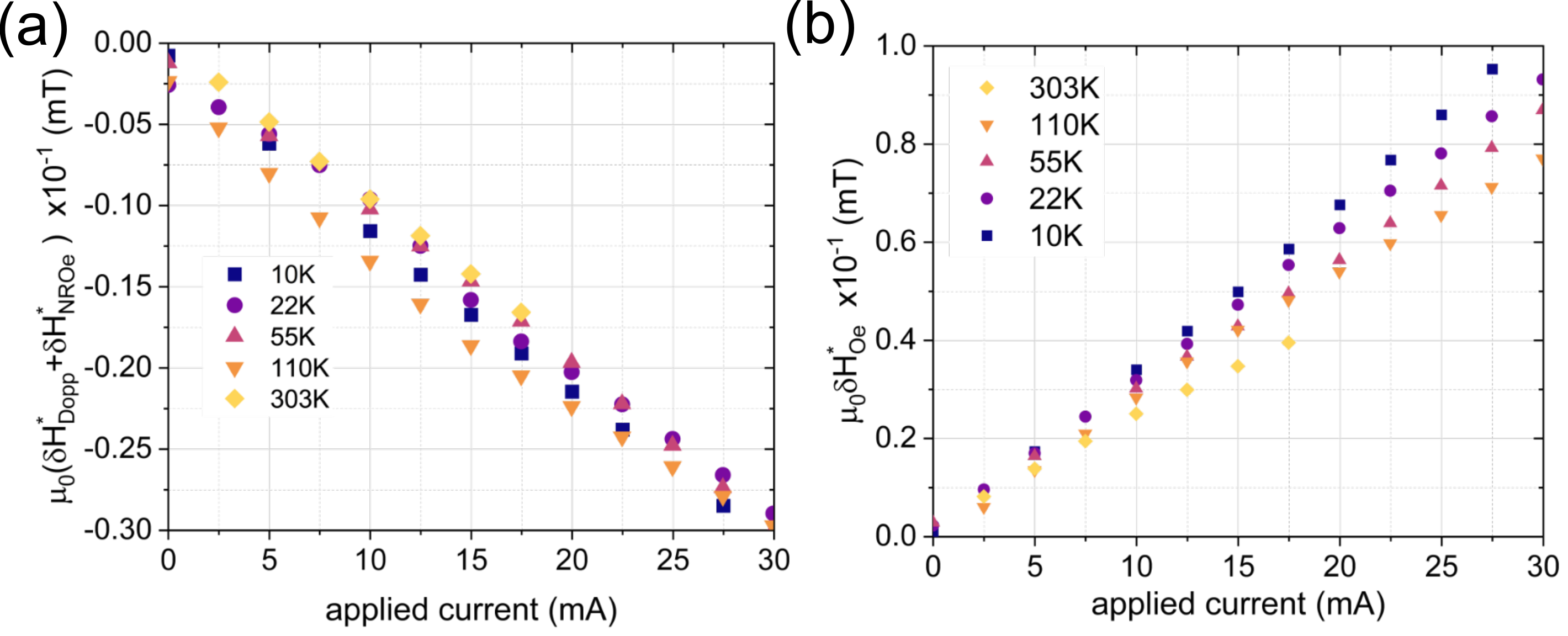}
\caption{Estimated current-induced field shifts in a Fe strip of width $w=13.92 \: \mu \text{m}$ and thickness $t$=20.7 nm, at a frequency of 19 GHz and wave number $k_1=3.9 \; \text{rad}/\mu \text{m}$. (a) Spin wave Doppler field and non-reciprocal Oersted shifts $\delta H_\text{Dopp}^* + \delta H^*_\text{NROe}$ as a function of applied current for different temperatures. (b) Oersted field shift $\delta H_\text{Oe}^*$ as a function of applied current for different temperatures. The data has been smoothed for clarity.} 
\label{fig:HShifts}
\end{figure}

We have measured the experimental shifts as explained in the main text. In Fig.~\ref{fig:HShifts} we present the field shifts $\delta H^*_\text{Dopp}+\delta H^*_\text{NROe}$ and $\delta H^*_\text{Oe}$. Note that a finite $\delta H^*_\text{Oe}$ means that there is an asymmetry in the distribution of the electric current through the cross section of the ferromagnetic strip \cite{HaidarThesis.2013}. From the sign of the shift we conclude that the electric current distribution is concentrated towards the bottom surface of the film. This could be caused by an increased electron-surface scattering at the top Fe/MgO interface due to the same atomic disorder that we have inferred from the observed spin wave field non-reciprocity [Fig.~\ref{fig:DKs}].  

\begin{figure}[H]
\centering
\includegraphics[width=0.85\textwidth]{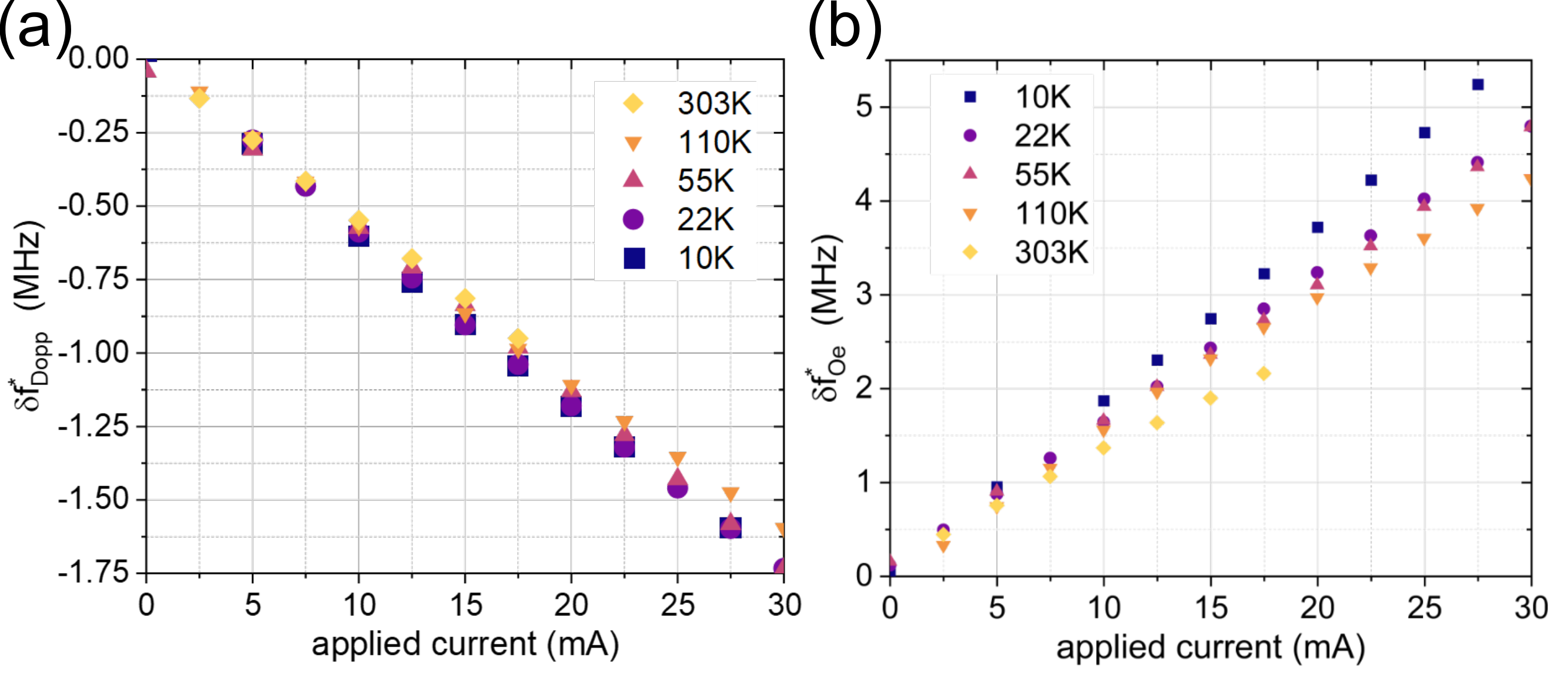}
\caption{Estimated current-induced frequency shifts obtained from the field shifts in Fig.~\ref{fig:HShifts}. (a) Spin wave Doppler frequency shift $\delta f_\text{Dopp}^*$ as a function of applied current for different temperatures after correcting for $\delta f^*_\text{NROe}$. (b) Oersted frequency shift $\delta f_\text{Oe}^*$ as a function of applied current for different temperatures.} 
\label{fig:FShifts}
\end{figure}

Now we can transform these field shifts into frequency shifts with $\delta f= \frac{\partial f}{\partial H} \delta H$ (Fig.~\ref{fig:FShifts}), where $\frac{\partial f}{\partial H} = - \frac{\partial f_0}{\partial H}$, and $f_0(H)$ is the resonance frequency as a function of the applied field. Note that the minus sign takes into account that for an experiment at fixed frequency $f_0$, a field shift $\delta H$ ($H\rightarrow H_0+\delta H$) corresponds to a frequency shift $\delta f$ such that $f_0=f(H_0) \rightarrow f(H_0 \pm |\delta H|) \approx f_0 \mp |\delta f|$.

To isolate $\delta f^*_\text{Dopp}$, we have estimated $\delta f^*_\text{NROe}$ using the analytical expression developed in reference \cite{Haidar.2014}. Indeed, both contributions have the same symmetry with respect to current and wave vector, and therefore cannot be separated with our measurements alone. 

%%%%%%%%%%%%%%%%%%%%%%%%%%%%%%%%%%%%%%%%%%%%%%%%%%%%%%%%%%%%%%%%%%%%%%%%%%%%%%%%%%%%%%%
\section{\label{sec:2CurrentModel}Details of the resistivity model\protect}

%%%%%%%%%%%%%%%%%%%%%%%%%%%%%%%%%
\subsection{Spin mixing resistivity}
The standard two-current model considers two parallel independent channels of conduction for the two spin directions. This model further assumes that the scattering events in each spin channel are independent of one another. To expand this model and lift this independence restriction, a possibility is to include spin-flip scattering with magnons. In such events, conduction electrons of one spin direction scatter with a magnon (flipping their spin) while retaining enough linear momentum to contribute significantly to the conduction of their new spin channel. In this way, this events effectively transfer current from one band to the other while creating additional resistivity to the overall transport. This additional resistivity contribution is denoted by the spin mixing resistivity $\rho_{\uparrow\downarrow}$. Under the two current model, this this term can be introduced as follows \cite{Fert.1969}:

%%%%%%%%%%%%%%%%
\begin{subequations}
\label{eq:transportMod}
\begin{eqnarray}
    \rho = 
   \frac{\rho_\uparrow \rho_\downarrow + \rho_{\uparrow\downarrow}(\rho_\uparrow + \rho_\downarrow)}{
   \rho_\uparrow + \rho_\downarrow + 4\rho_{\uparrow\downarrow}},,\label{eq:rhoMod}
\end{eqnarray}
\begin{equation}
    P=\frac{\rho_\downarrow-\rho_\uparrow }{\rho_\uparrow + \rho_\downarrow + 4\rho_{\uparrow\downarrow} }.\label{eq:polarizationMod}
\end{equation}
\end{subequations}
%%%%%%%%%%%%%%%%

\begin{figure}[H]
\centering
\includegraphics[width=0.45\textwidth]{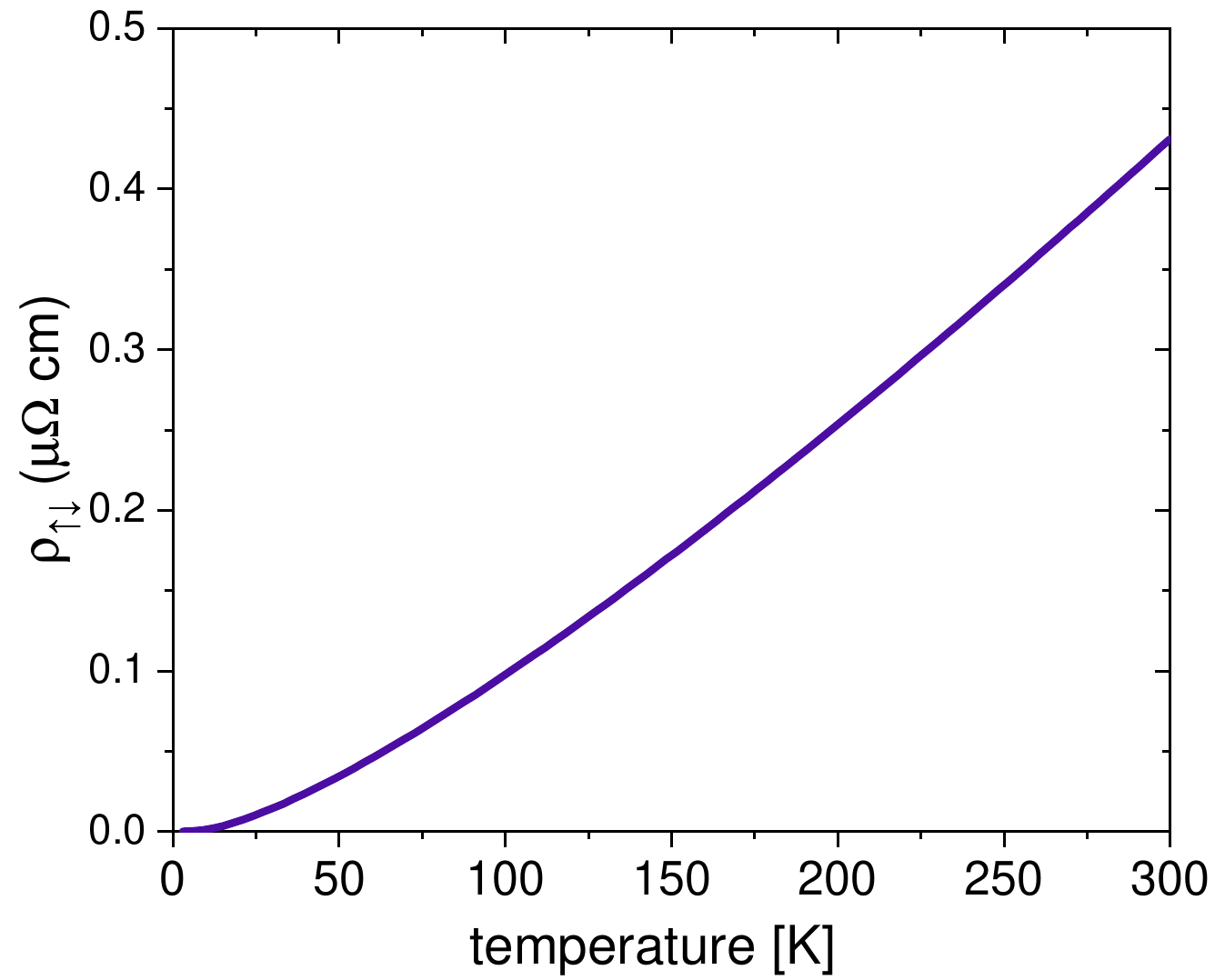}
\caption{Spin mixing resistivity contribution as a function of temperature.} 
\label{fig:Updown}
\end{figure}

To estimate the spin-mixing resistivity $\rho_{\uparrow\downarrow}$ we used the s-d electron-magnon scattering model and band structure parameters reported for Fe in ref.~\cite{Fert.1969}. In Fig.~\ref{fig:Updown} we plot the temperature dependence of $\rho_{\uparrow\downarrow}$. The values obtained from this model are much lower than all measured resistivities, consequently, introducing such spin-mixing contribution in the model does not modify significantly the estimations of $\rho_{\uparrow}$ and $\rho_{\downarrow}$. This explains why we did not include this contribution in our model.

Note that in our model, we have divided the electron-magnon spin-flip scattering in two parts: 1) events in which the conduction electrons of one spin direction scatter into conduction electrons of the opposite spin band, at the origin of $\rho_{\uparrow\downarrow}$; 2) events in which the conduction electrons of one spin direction scatter into heavy electron states of the opposite spin band, leading to $\rho_{mag,\text{i}}$. In Eqs.~\eqref{eq:transportMod}, we observe that the main role of the first type of events is to decrease the degree of spin polarization of the current through $\rho_{\uparrow\downarrow}$ as conduction electrons are transferred between the two channels. On the other hand, the second type of events can actually contribute to a spin-polarization through very different $\rho_{mag,\text{i}}$; this is what we have determined for Fe [Fig. 3 (c) in main text].

%%%%%%%%%%%%%%%%%%%%%%%%%%%%%%%%%
\subsection{Fuchs-Sondheimer model for residual resistivities}
At low temperatures the surfaces of the film should limit the conductivity. Under the Fuchs-Sondheimer model \cite{Zhou.2018,Sondheimer.1952} we can write the residual resistivities as $\rho_\text{o,i}= r_i \frac{\lambda_i}{t}$, where $r_i$ are spin dependent specularity coefficients, $\lambda_i$ are the spin dependent mean free paths and $t$ is the thickness of the film. 

At low temperature, where the resistivity contributions from thermal sources of scattering are negligible, the low temperature degree of spin-polarization should be given only by the electron-surface scattering $P(T \rightarrow 0K)\approx (r_\downarrow \lambda_\downarrow- r_\uparrow \lambda_\uparrow)/(r_\downarrow \lambda_\downarrow + r_\uparrow \lambda_\uparrow)$. 

The high $P$ we estimate experimentally means that the spin-up channel must be much more conductive than the spin-down one, then we can assume that $\lambda_\uparrow > \lambda_\downarrow$. At the same time, this requires a strongly spin-polarized specularity, i.e. $r_\uparrow \ll r_\downarrow$.

As we suggest in the main text, a strongly spin-polarized specularity of the Fe/MgO and MgO/Fe interfaces could occur if the interfaces are rough enough. Note that a roughness of a few atomic layers is enough to modify importantly the resistivity \cite{Timoshevskii.2008}. Indeed, we believe that the top interface of our Fe films present some atomic roughness as suggested by magnetic disorder at the top Fe/MgO interface [small perpendicular magnetic anisotropy in Fig.~\ref{fig:DKs} (c)] and concentration of electric current towards the bottom interface [large and positive $\delta f_\text{Oe}^*$ in Fig.~\ref{fig:FShifts} (b)]. Finally, we argue that such roughness can lead to a spin selectivity at the Fe/MgO, as spin up electron wave functions can penetrate further into the MgO \cite{Bowen.2001} making these states less sensitive to scattering at the interface. This selectivity would be at the microscopic origin of the high spin-polarization of the residual resistivities we observe experimentally. 

%%%%%%%%%%%%%%%%%%%%%%%%%%%%%%%%%
\subsection{Electron-magnon scattering model}
The model we have used to describe the temperature dependence of the electron-magnon scattering is based on the model developed by Goodings \cite{Goodings.1963}: highly conductive "s" spin up (down) electrons are scattered into heavy "d" spin down (up) states by magnons through spin-flip events. Raquet et al. \cite{Raquet.2002} extended this model to include the external magnetic field (Zeeman term) and temperature dependence into the spin wave energy (renormalization of the spin wave stiffness $D$). The result is an expression for the total resistivity $\rho^*_{mag}(H,T)$ due to the electron s-d scattering mediated by magnons as a function of the temperature and the applied magnetic field $H$. 

This model can be used to interpret longitudinal magnetoresistance measurements if one assumes that the main effect of a magnetic field is to modify the magnon populations and therefore modify the effective electron-magnon scattering. In this way, the longitudinal magnetoresistance becomes a direct measurement of the change of resistivity due to electron-magnon scattering:

%%%%%%%%%%%%%%%%
\begin{equation}
\begin{aligned}
    \label{eq:MMR}
   \Delta \rho(H,T_\text{o}) = \rho(H,T_\text{o})-\rho(0,T_\text{o})=\rho^*_{mag}(H,T_\text{o})-\rho^*_{mag}(0,T_\text{o}).
\end{aligned}
\end{equation}
%%%%%%%%%%%%%%%%

By fitting magnetoresistance data to the model proposed by Raquet et al. \cite{Raquet.2002} one can obtain the spin wave stiffness $D(T_\text{o})$ at given temperature $T_\text{o}$ as a fitting parameter. After we obtain $D$ for a sufficiently large set of temperatures, we can fit it to a renormalization function of the type \cite{Stringfellow.1968}

%%%%%%%%%%%%%%%%
\begin{equation}
\begin{aligned}
    \label{eq:Renormalization}
   D(T)=D_0\left(1-d_1T^2-d_2T^{5/2}\right).
\end{aligned}
\end{equation}
%%%%%%%%%%%%%%%%

In our case we can only reliably fit up to the second order term: $D(T)=356[1-(2\times10^{-6})T^2]$ $\text{meV}\text{\AA}^2$. We can compare this with the result obtained from FMR: $D_\text{FMR}=268[1-(1.8\times10^{-6})T^2+(4\times10^{-8})T^{5/2}]$ $\text{meV}\text{\AA}^2$ [see Fig.~\ref{fig:FMR} (b)]. To calculate the spin wave stiffness we have used $D_\text{FMR}=2g\mu_\text{B}A/M_\text{o}$ ($g$ is the g-factor, $\mu_\text{B}$ the Bohr magneton and $M_\text{o}$ the saturation magnetization at 0 K). The scaling difference between the two methods could be attributed to the two types of magnons probed in each experiment ($k \approx 0 $ for FMR, and large $k$ magnons for the magnetoresistance measurements). 

Finally, to reconstruct the temperature dependence of the resistivity due to electron-magnon scattering [$\rho^*_{mag}(H,T)$], we plug the obtained $D(T)$ back into the model. The result is plotted in the inset of Fig. 4 of the main text for an applied field $\mu_0 H$=120 mT. Note that we have used the same Fe band structure parameters as those used in ref.~\cite{Raquet.2002}.

\subsection{Two-current electron-magnon scattering}
 
Although the model proposed by Raquet et al. \cite{Raquet.2002} considers spin-flip scattering events, for simplicity, it does not make the spin-polarized two-current separation in its standard formulation. It can be adapted to be two-current compatible if one introduces spin-polarized band structure parameters (density of states, exchange coupling constants, etc.). As these parameters are not known and the equations become rather cumbersome, we have taken an empirical minimalist approach for its two-current generalization: we have introduced renormalizing factors $B_i$ ($i=\uparrow, \downarrow$) that become $"$weights$"$ for each spin channel that take into account the net effect from the spin-polarized band structure parameters:

%%%%%%%%%%%%%%%%
\begin{equation}
\begin{aligned}
    \label{eq:SPElectronMagnon}
   \rho_{mag,i}(H,T)=B_i\rho^*_{mag}(H,T).
\end{aligned}
\end{equation}
%%%%%%%%%%%%%%%%

\FloatBarrier
%%%%%%%%%%%%%%%%%%%%%%%%%%%%%%%%%%%%%%%%%%%%%%%%%%%%%%%%%%%%%%%%%%%%%%%%%%%%%%%%%%%%%%%%%
%\nocite{*}
\bibliography{Supplemental}% Produces the bibliography via BibTeX.
\bibliographystyle{apsrev4-2}